\documentclass[superscriptaddress, twocolumn, prb, showpacs]{revtex4} 

\usepackage{amssymb}
\usepackage{hyperref}
\usepackage{amsmath}
\usepackage{bbm}
\usepackage{graphicx}
\usepackage{natbib}

\begin{document}

\author{Thomas Ayral}
\affiliation{Theoretische Physik, ETH Zurich, 8093 Z{\"u}rich, Switzerland}
\affiliation{Centre de Physique Th{\'e}orique, Ecole Polytechnique, CNRS-UMR7644, 91128 Palaiseau, France}

\affiliation{Institut de Physique Th\'eorique (IPhT), CEA, CNRS, URA 2306, 91191 Gif-sur-Yvette, France}

\author{Silke Biermann}
\affiliation{Centre de Physique Th{\'e}orique, Ecole Polytechnique, CNRS-UMR7644, 91128 Palaiseau, France}
\affiliation{Japan Science and Technology Agency, CREST, Kawaguchi
  332-0012, Japan}
\author{Philipp Werner}
\affiliation{Department of Physics, University of Fribourg, 1700 Fribourg, Switzerland}
\affiliation{Theoretische Physik, ETH Zurich, 8093 Z{\"u}rich, Switzerland}

\title{

Screening and Non-local Correlations 
in the Extended Hubbard Model from
Self-Consistent Combined GW and Dynamical
Mean Field Theory

}

\date{today}

\begin{abstract}
We describe a recent implementation of the combined GW and dynamical
mean field (DMFT) method ``GW+DMFT'' for the
two-dimensional Hubbard model with on-site and nearest-neighbor repulsion. 
We clarify the relation of the GW+DMFT scheme to alternative approaches
in the literature, and discuss the corresponding approximations to the
free energy functional of the model.
Furthermore, we describe a numerically exact technique for the
self-consistent solution of the GW+DMFT equations, namely the
hybridization expansion continuous-time algorithm for solving 
the dynamical impurity model that arises within the GW+DMFT scheme.
We compute the low-temperature phase diagram of the extended
Hubbard model, addressing the metal-insulator transition
at small intersite interactions and the transition to a
charge-ordered state for stronger intersite repulsions.
Within the GW+DMFT framework, as in extended DMFT, the intersite
repulsion translates into a frequency-dependence of the local
effective interaction. We analyze this dependence by extracting
a charateristic plasma frequency and show how it affects
the local spectral function. 

\end{abstract}

\pacs{ 71.10.Fd}

\maketitle

\section{Introduction}

Understanding the effects of strong electronic correlations
in lattice systems remains a challenge in condensed matter
physics. The competition between the tendency of electrons to delocalize
due to the corresponding gain in kinetic energy and 
localisation by the Coulomb interaction gives rise to a panoply
of interesting phenomena, ranging from simple mass enhancements
in the sense of Landau theory to charge-, spin- or orbital ordering
phenomena.
To reduce the complexity of the problem, while still keeping the
main qualitative effects, {\it e.~g.}, of the delocalisation-localisation
transition, one often resorts to low-energy effective models
such as the Hubbard or Anderson lattice models.
The two-dimensional single-band Hubbard Hamiltonian
with a static on-site repulsive interaction $U$ for example
is believed to capture the physics of the high-temperature 
superconducting cuprates.\cite{Zhang_1989}
Charge-ordering transitions can be captured when also an
intersite interaction term, mimicking the longer range
Coulomb interactions, is retained. 
This motivates the construction of the extended Hubbard model,
where charge-ordering effects and screening of the local
interactions due to the non-local ones are included in 
addition to the pure Hubbard model physics.

In the paramagnetic phase at half-filling,
the Hubbard model exhibits a Mott transition from a 
metal to a Mott insulator whose spectral function is 
characterized by a gap at the Fermi energy 
and Hubbard bands corresponding to atomic-like excitations. 
This behavior is captured by computational schemes such as 
the Dynamical Mean Field Theory (DMFT) (see Refs.~\onlinecite{Georges_1996}  and \onlinecite{Kotliar_Vollhardt_2004} for review)
or its extensions (C-DMFT\cite{Lichtenstein_2000,Kotliar_2001}, DCA\cite{Hettler_1998}, dual fermions\cite{Rubtsov_2008}, dual bosons\cite{Rubtsov_2012}, $D\Gamma A$\cite{Toschi_2007}, DMFT+$\Sigma_k$\cite{Kuchinskii_2005}).
A formalism which allows to treat screening by non-local 
interactions is extended DMFT (EDMFT).\cite{Si_1996,Sengupta_1995, Kajueter_thesis,Sun_2002,Sun_2004}
Its combination with the ab-initio GW 
approach \cite{Biermann_2003,Aryasetiawan_Biermann_2004,biermann-gwdmft-arxiv0401653}
introduces some momentum-dependence also into the 
self-energy, capturing 
the interplay of screening and non-local correlations. 
This scheme allows for a self-consistent computation of 
the effective ``Hubbard $U$" in a solid and a fully parameter-free 
ab-initio simulation approach for correlated materials.
The idea is to take the local part of the self-energy from the 
EDMFT calculation and supplement it by the nonlocal component of the 
GW self-energy. A rigorous functional formulation -- which is 
detailed in section II -- puts this theory on the same level
of mathematical rigor as, {\it e.~g.}, the functional formulation of 
Hohenberg-Kohn density functional theory.

Despite these promising perspectives, the technical difficulties
associated with the numerical treatment of the frequency-dependent 
effective interaction have so far prevented a self-consistent
calculation of spectral properties within GW+DMFT, even at the model
level. Recent progress, both within approximate schemes 
\cite{Casula_2012} and numerically exact Monte Carlo techniques,
\cite{Werner_2010} is currently giving a new impact to the
field.\cite{Werner_2012, Casula_effmodel_2012, Ayral_2012}
In particular, the development of efficient continuous-time
Monte Carlo techniques,\cite{Werner_2007, Rubtsov_2005} generalized
to dynamical interactions in Ref.~\onlinecite{Werner_2010}, 
now allows for the fully self-consistent solution of the
GW+DMFT equations, with high enough accuracy to also extract
spectral properties.

In this paper we use these state-of-the-art techniques to study an extended Hubbard model with on-site
interaction $U$ and nearest-neighbor repulsive interaction $V$, and explore the 
interplay of screening and non-local correlations. We show that this model, 
if solved within the EDMFT framework, captures
dynamical screening effects related to the nonlocal interaction $V$, 
high-energy satellite features in the one-particle spectra and,
for large $V$, a transition to a charge-ordered state. 
We then proceed to study how diagrammatic corrections to EDMFT in
the form of momentum-dependent GW contributions to the self-energy modify
this picture, and compare the results of self-consistent 
and non-self-consistent 
implementations.

The paper is organized as follows: 
in Section II, we discuss the model. In particular,
we will show how the Hamiltonian formulation of the problem can be
recast into an action or functional formulation.
We will explicitly construct two flavors of free energy functionals 
whose stationary points would give the exact solution of the model,
the first one due to Almbladh {\it et al.}\cite{Almbladh_1999} (see also Ref. \onlinecite{Chitra_2001}), the second one
constructed by Sun {\it et al.}.\cite{Sun_2002} 
In section III, we present different methods of solution: extended
DMFT, GW and combined schemes. We argue that the GW+DMFT formalism 
can be derived from the Almbladh free energy functional, 
and discuss the differences to the scheme proposed in Ref. \onlinecite{Sun_2002}. The latter stems from another energy functional, 
which would correspond to a ``GD+SOPT+DMFT'' method, where $D$ is the
boson propagator associated to screening of the non-local interaction
only, used in a GW-like fashion, supplemented by second order
perturbation theory (SOPT) for the non-local effects of the
local interaction and dynamical mean field theory.
Numerical techniques for the solution of the equations are
described in Section IV and the computational scheme used in the present work is
summarized in Section V. The results of our study are presented in Section VI.
Section VII contains a summary of our most
important findings, and provides perspectives how our work inserts
itself into the field.

\section{Model}
\subsection{Hamiltonian Formulation}

We consider the single-band $U$-$V$ Hubbard model on a two-dimensional square lattice,
defined by the grand-canonical Hamiltonian
\begin{eqnarray}
H&=&-t\sum_{\langle ij\rangle\sigma}\left(c_{i\sigma}^{\dagger}c_{j\sigma}+h.c\right)-\mu\sum_{i}n_{i}\nonumber\\
&+&U\sum_{i}n_{i\uparrow}n_{i\downarrow}+
V \sum_{\langle ij\rangle}n_{i}n_{j},
\label{eq:U-V_model}
\end{eqnarray}
where $c_{i\sigma}$ and $c_{i\sigma}^{\dagger}$ denote the annihilation
and creation operators of a particle of spin $\sigma=\uparrow,\downarrow$
at the lattice site $i$, $n_{i\sigma}=c_{i\sigma}^{\dagger}c_{i\sigma}$, and
$n_{i}=n_{i\uparrow}+n_{i\downarrow}$. $\sum_{\langle ij\rangle}$ denotes the sum over nearest-neighbor bonds, $t>0$ is the hopping integral between two neighboring
sites, $\mu$ is the chemical potential, $U$ the on-site interaction
between electrons of opposite spin and $V$ the interaction between
two electrons on neighboring sites, irrespective of their spin. The number of
nearest neighbors is $z=2d=4$, where $d$ is the dimension.

This model can be obtained from first principles in realistic calculations, as discussed in Appendix \ref{remarks_extended_hubbard_models}. The limiting case $V=0$ corresponds to the conventional Hubbard
model.\cite{Gutzwiller_1963,Hubbard_1963,Kanamori_1963} In the present study, we will limit ourselves
to repulsive interactions, $U>0$ and $V>0$.

In the limit of large $V$, the extended Hubbard model
has been shown to display a transition to a charge-ordered state characterized
by a freezing of charge carriers and a spatial
modulation of the charge density (charge-density wave, CDW). This
can be explained by a simple energetical argument in the strongly correlated
regime ($U\gg t$): while for $U/V\gg1$, electrons will lower their
energy by arranging themselves one per site to minimize the on-site
repulsion, for $V/U\gg1$, electrons will minimize their off-site
repulsion by choosing an arrangement such that one sublattice is occupied
by two electrons per site while the other is empty, leading to a commensurate
charge order. In the metallic phase ($U\ll t$), the effect of $V$
is more easily understood in terms of screening: the charge fluctuations
induced by the $V$ term lead to a reduction of the local effective
interaction.

The $U$-$V$ Hubbard model has been studied by a variety of approximate methods.
An early study in the zero-overlap limit ($U/t\gg1$) has predicted
a phase transition between a Mott insulator and a charge-ordered (CO)
insulator at $V_{c}=U/z$ at zero temperature,\cite{Bari_1971} while
weak-coupling mean-field studies have predicted a transition between
antiferromagnetic order (AFM, \emph{i.e.} commensurate SDW) and charge
order (\emph{i.e.} commensurate CDW) at the same boundary.\cite{Wolff_1983,Yan_1993}
This has been confirmed by Monte-Carlo calculations in two dimensions.\cite{Zhang_Callaway_1989}
The $V_{c}=U/z$ boundary has been shown to hold in the $U/V\rightarrow0$ limit
as well as in the $U/V\rightarrow\infty$ limit by a study at half-filling
in the infinite dimensional limit.\cite{vanDongen_1991} Higher-order
corrections have been considered in Refs.~\onlinecite{vanDongen_1994} and \onlinecite{vanDongen_1995},
leading to a renormalization of the critical temperature and order
parameter, as well as the prediction of phase separation in the zero-temperature
limit. More recently, variational cluster \cite{Aichhorn_2004} and
two-particle self-consistent approaches \cite{Davoudi_2007}
have been applied to the $U$-$V$ Hubbard model.

A first DMFT treatment described the opening of a {}``pseudo-gap''
and the reentrant behavior of the critical $V_{c}$ as a function
of temperature.\cite{Pietig_1999} In this simple scheme the $V$-term
only contributed at the Hartree level by shifting the chemical potential,
since in the limit of infinite dimensions, the contributions beyond
Hartree of non-local terms vanish, while local terms such as the on-site
Hubbard $U$ do not vanish under rescaling.\cite{Muller_1989}
The screening effects contained in the $V$-term are
not captured by standard single-site DMFT. Therefore, an extended DMFT (EDMFT)
scheme has been proposed to remedy this shortcoming.\cite{Si_1996,Sengupta_1995, Kajueter_thesis,Sun_2002,Sun_2004}
Within this scheme, the non-local interactions induce a frequency-dependence
of the effective local interaction and lead to a sizeable \emph{reduction}
of the static value of $U$. In addition, Refs.~\onlinecite{Sun_2002} and \onlinecite{Sun_2004} 
showed that one of the consequences of adding a spatially non-local 
contribution to the self-energy is to make the system more
insulating. However, none of the previous studies has given a precise account of how $U$ and
$V$ affect the properties of the local frequency-dependent interactions
nor how the latter in turn modify the spectral properties of the system.
A systematic investigation of the effect of non-local GW contributions 
to the self-energy is also lacking.

In the present work, we try to clarify these issues, restricting ourselves to the 
paramagnetic phase at half-filling. We will give all energies in units of the half-bandwidth, $D=4t$. The
inverse temperature will be denoted by $\beta=1/T$ ($k_{B}=1$).

\subsection{Action Formulation}

The solution of model~(\ref{eq:U-V_model}) amounts to computing the
Green functions and other correlation functions. For this purpose, 
it is convenient to write the grand-canonical partition function $Z=\mathrm{Tr}e^{-\beta H}$
as a coherent-state path integral\cite{Negele_1988} $Z=\int\mathcal{D}[c_{i}^{*},c_{i}]e^{-S}$ where
\begin{eqnarray}
&&S[c^{*},c]=\int_{0}^{\beta}d\tau\left\{ \sum_{ij\sigma}c_{i\sigma}^{*}(\tau)\left(\left(\partial_{\tau}-\mu\right)\delta_{ij}+t_{ij}\right)c_{j\sigma}(\tau)\right.\nonumber\\
&&\left.+U\sum_{i}n_{i\uparrow}(\tau)n_{i\downarrow}(\tau)+\frac{1}{2}\sum_{ij} v^{nl}_{ij}n_{i}(\tau)n_{j}(\tau)\right\}. \label{eq:U-V_action}
\end{eqnarray}
\and $c_{i}^{*}$ and $c_{i}$ denote conjugate anticommuting Grassmann fields for site $i$ ($c_i(\tau+\beta)=-c_i(\tau)$), $t_{ij}=-t\delta_{\langle ij\rangle}$, $v^{nl}_{ij}=V\delta_{\langle ij\rangle}$ and $\delta_{\langle ij\rangle}=1$
if $i$ and $j$ are nearest neighbors and 0 otherwise. $\tau$ is the
imaginary time variable. We will denote by $\omega_{n}=(2n+1)\pi/\beta$
($\nu_{n}=2n\pi/\beta$) the corresponding fermionic (bosonic) Matsubara
frequencies.  The Fourier transforms
of $t_{ij}$ and $v_{ij}^{nl}$ on the lattice are:
\begin{eqnarray}
\epsilon_{k}&=&-2t\left(\cos(k_{x})+\cos(k_{y})\right),\\
v^{nl}_{k}&=&2V\left(\cos(k_{x})+\cos(k_{y})\right).
\end{eqnarray} 

Using the identity $n_i n_i = 2 n_{i\uparrow}n_{i\downarrow}+n_i$, we can rewrite the interaction terms of Eq.~(\ref{eq:U-V_action}) as 
$\frac{1}{2}\sum_{ij} v_{ij}n_{i}(\tau)n_{j}(\tau)$ with $v_{ij}=U\delta_{ij}+V\delta_{\langle ij \rangle}$,  or 
\begin{equation}
v_k=U+v^{nl}_k,
\end{equation}
provided that we shift the chemical potential $\mu\rightarrow \tilde \mu = \mu +U/2$. The action hence becomes
\begin{eqnarray}
&&S[c^{*},c]=\int_{0}^{\beta}d\tau\left\{ \sum_{ij\sigma}c_{i\sigma}^{*}(\tau)\left(\left(\partial_{\tau}-\tilde \mu\right)\delta_{ij}+t_{ij}\right)c_{j\sigma}(\tau)\right.\nonumber\\
&&\left.+\frac{1}{2}\sum_{ij} v_{ij}n_{i}(\tau)n_{j}(\tau)\right\}. \label{eq:tilde_V_action}
\end{eqnarray}

\subsection{Functional Formulation}

The problem of finding the solution to the Hamiltonian model
(\ref{eq:U-V_model}) or calculating the Green's function corresponding 
to the action (\ref{eq:U-V_action}) can equivalently be formulated
in a functional language.
The familiar Luttinger-Ward or Baym-Kadanoff functionals provide
examples of such a construction. In the present context, a
formulation in terms of the free energy written as a
functional of both, the Green's function $G$ and the screened 
Coulomb interaction $W$, is the method of choice, since the combined 
GW+DMFT method can naturally be viewed
as a specific approximation to such a functional.
Indeed, the GW+DMFT solution as formulated in Refs.~\onlinecite{Biermann_2003} and \onlinecite{
Aryasetiawan_Biermann_2004} can be derived as a stationary point $(G, W)$
of the free energy functional introduced by Almbladh {\it et al.},\cite{Almbladh_1999} after approximating the correlation part of
this functional by a combination of local and non-local terms
stemming from DMFT and GW, respectively.
To draw an illustrative analogy, GW+DMFT provides an approximation
to the correlation part of the Almbladh free energy functional, 
on the same footing as the local density approximation of density
functional theory \cite{Kohn_1999} is 
an approximation to the exchange-correlation part of the
Hohenberg-Kohn functional of the energy.

In the literature, several variants of functionals of $G$ and
$W$ have been discussed,\cite{Chitra_2001, Sun_2002, Aryasetiawan_Biermann_2004}
and different derivations given.
Using a Hubbard-Stratonovich (HS) decoupling as 
in Ref.~\onlinecite{Biermann_2003}, we discuss two flavors of free
energy functionals which differ by the choice of the part of 
the interaction that is decoupled by the HS transformation.
The first one reproduces the $\Psi$ functional introduced
by Almbladh on which the GW+DMFT construction of Ref.~\onlinecite{Biermann_2003}
is based. The second one is a variant that was used in the
study of the extended Hubbard model in Ref.~\onlinecite{Sun_2002}.

The Hubbard-Stratonovich transformation \cite{Hubbard_1959} 
relies on the following identity:
\begin{eqnarray}
\exp\left( \frac{1}{2}\int_0^{\beta} d\tau b_i(\tau) A_{ij} b_j(\tau)\right)=\int \frac{\mathcal{D}[x_1(\tau),x_2(\tau),\dots]}{\sqrt{(2\pi)^N \det A	}}\nonumber\\
\times \exp\left(\int_0^{\beta} d\tau \left\{ -\frac{1}{2}x_i(\tau)[A^{-1}]_{ij}x_j(\tau)\mp x_i(\tau) b_i(\tau)\right\}\right)
\label{eq:HS}
\end{eqnarray}
where $A$ is a real symmetric positive-definite matrix, $b_i(\tau)$ and $x_i(\tau)$ are periodic real fields ($b_i(\tau+\beta)=b_i(\tau)$), and summation over repeated indices is understood. In the following, we will choose the upper sign for the last term.

Decoupling the whole (local and non-local) interaction 
$\frac{1}{2}\sum_{ij} v_{ij}n_{i}(\tau)n_{j}(\tau)$ 
by the HS transformation corresponds to applying the above
formula (\ref{eq:HS}) to the interaction term in 
Eq.~(\ref{eq:tilde_V_action}), that is to the choice
$b_i\equiv i n_i$, $A_{ij}\equiv v_{ij}$ and $x_i\equiv \phi_i$.
This choice -- denoted as HS-UV in the following -- leads to the 
construction of the $\Psi$ functional as in Almbladh {\it et al.}\cite{Almbladh_1999} 
and the formalism of Ref.~\onlinecite{Aryasetiawan_Biermann_2004}. 
Within the approximation introduced in the next section this
corresponds to the GW+DMFT scheme as introduced in Ref.~\onlinecite{Biermann_2003}.
This approach (which was also used in Ref.~\onlinecite{Sun_2004}) 
relies on the argument that the two terms, which represent different 
matrix elements of the same interaction, should be treated on the same 
footing. 

In Ref.~\onlinecite{Sun_2002}, on the other hand,
the HS transform has been applied only to the non-local interaction term 
$\frac{1}{2}\sum_{ij} v^{nl}_{ij}n_{i}(\tau)n_{j}(\tau)$ 
in the action (\ref{eq:U-V_action}). This approach, dubbed
HS-V in the following, leads to a modified free energy functional,
which we denote by $\Psi_V$.
It was motivated by the aforementioned fact that in the limit of 
infinite dimensions,
with the non-local interaction rescaled as $V\rightarrow V/z$,
 the non-local term results in a trivial shift 
of the chemical potential, while the on-site interaction remains 
non-trivial, justifying a separate treatment for the non-local term. 

We will first explicitly work out the construction
of the two functionals $\Psi$ and $\Psi_V$ and in section~\ref{results},
we will compare the results from both decoupling strategies. 
The results in section \ref{screening}, finally, are based on the HS-UV decoupling, 
that is, on the $\Psi$ functional 
and the GW+DMFT formalism of Ref.~\onlinecite{Biermann_2003}.

\subsubsection{``UV-Decoupling'': the $\Psi$ Functional}

In the HS-UV decoupling scheme, the full interaction term is decoupled 
via an auxiliary bosonic field $\phi_{i}$. Choosing $b_i\equiv i n_i$, 
$A_{ij}\equiv v_{ij}$ and $x_i\equiv \phi_i$, the transformation 
(\ref{eq:HS}) applied to the action (\ref{eq:U-V_action}) leads to
\begin{align}
&S[c^{*},c,\phi] = \int_{0}^{\beta}d\tau\Bigg\{ -\sum_{ij\sigma}c_{i\sigma}^{*}(\tau)\left[(G_{0}^H)^ {-1}\right]_{ij}c_{j\sigma}(\tau) \Bigg\}  \nonumber\\
&+ \int_{0}^{\beta}d\tau\Bigg\{ \frac{1}{2}\sum_{ij}\phi_{i}(\tau)[v^{-1}]_{ij}\phi_{j}(\tau)+i \alpha 
\sum_{i}\phi_{i}(\tau)n_{i}(\tau)\Bigg\}, 
\end{align}
where we introduced the 
fermionic lattice Hartree Green's 
function $G^H_0$ defined by 
$\left[G_{0}^{H -1}\right]_{ij}\equiv((-\partial_{\tau}+\mu
+\frac{U}{2})\delta_{ij}-t_{ij})$.
For later use, we have moreover inserted a coupling constant
$\alpha$ in front of the fermion-boson coupling term.
The physically relevant case corresponds to $\alpha$=1.

The HS transformation replaces an electron-electron interaction by 
an electron-boson interaction, and introduces a new variable, the 
auxiliary real boson field $\phi$. This has an important consequence: 
even a first-order diagram in this new interaction contains diagrams 
of infinite order in the electron-electron interaction. Making 
even simple approximations on the new action can thus lead to 
non-trivial diagrams for the original action. Moreover, the 
electron-boson vertex $i\phi_{i}n_{i}$ is \emph{local}. This locality 
ensures that in the limit of infinite dimensions, the interactions 
(and in particular $V$) will contribute beyond the Hartree level. 

The generating functional of correlation functions is obtained by
introducing the bilinear sources $J_{\mathrm{f}}$ and $J_{\mathrm{b}}$, coupling to
the fermionic and bosonic operators respectively, so that the action 
becomes $S[c^{*},c,\phi] - S[J_{\mathrm{f}},J_{\mathrm{b}}]$, with
\begin{eqnarray}
S[J_{\mathrm{f}},J_{\mathrm{b}}]=\sum_{ij}\int_{0}^{\beta}d\tau d\tau'\bigg(J_{\mathrm{f},ij}(\tau,\tau')c_{i}^{*}(\tau)c_{j}(\tau')\nonumber\\
+\frac{1}{2}J_{\mathrm{b},ij}(\tau,\tau')\phi_{i}(\tau)\phi_{j}(\tau')\bigg).
\end{eqnarray}
The fermionic and bosonic Green's functions for this action are $G_{ij}(\tau-\tau')=-\langle Tc_{i}(\tau)c_{j}^{\dagger}(\tau')\rangle=\delta\Omega/\delta J_{\mathrm{f},ij}(\tau,\tau')$
and $W_{ij}(\tau-\tau')=\langle T\phi_{i}(\tau)\phi_{j}(\tau')\rangle=-2\delta\Omega/\delta J_{\mathrm{b},ij}(\tau,\tau')$, where we have defined 
\begin{equation}
\Omega \equiv -\ln Z[J_{\mathrm{f}},J_{\mathrm{b}}]
=-\ln \mathrm{Tr} e^{-S[c^{*},c,\phi]+S[J_{\mathrm{f}},J_{\mathrm{b}}]}.
\label{Z}
\end{equation}

The non-interacting Green's functions (obtained by setting 
$i\phi_{i}n_{i}=0$) are respectively $G|_{\alpha=0}=G^H_{0}$ and
$W(k,i\nu_{n})|_{\alpha=0}=\left[ v_{k}^{-1}\right]^{-1} = v_{k}$. 
This gives
a first hint as to the physical meaning of $W$: without renormalization by 
the auxiliary bosons, it corresponds to the bare interaction. 
Coupling to the bosons, which represent density fluctuations of the
system, explicitly introduces screening into the physical description.

We next perform a Legendre transformation with respect to the sources 
$J_{\mathrm{f}}$ and $J_{\mathrm{b}}$, 
\begin{equation}
\Gamma[G,W]=\Omega[J_{\mathrm{f}}[G],J_{\mathrm{b}}[W]]-\mathrm{Tr}J_{\mathrm{f}}G+\frac{1}{2}\mathrm{Tr}J_{\mathrm{b}}W
\end{equation}
 with the reciprocity relations $J_{\mathrm{f}}=-\frac{\delta\Gamma}{\delta G}$
and $J_{\mathrm{b}}=2\frac{\delta\Gamma}{\delta W}$. The physical Green's functions will be obtained by setting $J_{\mathrm{f}}=0$ and $J_{\mathrm{b}}=0$, or equivalently,
by requiring the stationarity of $\Gamma$ with respect to $G$ and
$W$.

The free energy functional $\Gamma$ can be written as 
\begin{equation}
\Gamma_{\alpha=1}=\Gamma_{\alpha=0}+\Psi,
\end{equation}
where we have defined 
\begin{equation}
\Psi\equiv\int_{0}^{1}d\alpha\frac{d\Gamma}{d\alpha}.
\end{equation}

$\Gamma$ is the well-known Baym-Kadanoff functional,\cite{Baym_1962} 
while $\Psi$ is the extension of the Luttinger-Ward functional 
$\Phi[G]$ to one- and two-particle propagators.\cite{Luttinger_1960}

The non-interacting ($\alpha=0$) part of the $\Gamma$-functional
is readily evaluated as 
\begin{eqnarray}
&&\Gamma_{\alpha=0}=\mathrm{Tr}\ln (-G)-\mathrm{Tr}(G_{0}^{-1}-G^{-1})G \nonumber\\
&&\hspace{12mm}-\frac{1}{2}\mathrm{Tr}\ln W+\frac{1}{2}\mathrm{Tr}(v^{-1}-W^{-1})W.\label{eq:BK_alpha_0_HSUV}
\end{eqnarray}
Indeed, when $\alpha=0$, the action becomes Gaussian and thus explicitly integrable, namely: $\Omega_{\alpha=0}=-\ln\mathrm{Det}[-G_{0}^{-1}+J_{\mathrm{f}}]-\ln(\mathrm{Det}[v^{-1}-J_{\mathrm{b}}])^{1/2}$.
The above definition $G=\delta \Omega/\delta J_{\mathrm{f}}$ imposes $(G_{0}^{-1}-J_{\mathrm{f}})G=1$
and similarly $(v^{-1}-J_{\mathrm{b}})W=1$ yielding Eq.~(\ref{eq:BK_alpha_0_HSUV}).
Finally, stationarity of the full $\Gamma$ reads $\frac{\delta\Gamma}{\delta G}=0=\frac{\delta\Gamma_{\alpha=0}}{\delta G}+\frac{\delta\Psi}{\delta G}=G^{-1}-G_{0}^{-1}+\frac{\delta\Psi}{\delta G}$
for $G$ and $0=-\frac{1}{2}(W^{-1}-v^{-1})+\frac{\delta \Psi} { \delta W}$ for $W$. Defining the self energies
as
\begin{eqnarray}
&&\Sigma = \frac{\delta\Psi}{\delta G}, \hspace{5mm} \Pi =  -2\frac{\delta\Psi}{\delta W},
\end{eqnarray}
yields Dyson's equations for $G$ and $W$:
\begin{eqnarray}
&&G^{-1}=G_{0}^{-1}-\Sigma, \hspace{5mm} W^{-1}=v^{-1}-\Pi.
\end{eqnarray}
Being ``$\Psi$-derivable", these self-energies
will obey conservation rules.\cite{Baym_1961}

The above formulation shows that, formally, solving the lattice problem 
defined by Eq.~(\ref{eq:U-V_action_decoupled}) amounts to evaluating 
the corresponding $\Psi$-functional, from which $\Sigma$ and $\Pi$, and 
in turn $G$ and $W$ can be derived. In section \ref{Methods-of-Solution}, 
we will describe two complementary ways of approximating this functional, 
EDMFT and GW, before showing how to merge the two approaches, thus arriving 
at the GW+DMFT free energy functional.

\subsubsection{``V-Decoupling'':  The $\Psi_{V}$ Functional}

In the HS-V scheme, proposed in Ref.~\onlinecite{Sun_2002}, only the non-local 
interaction term is decoupled via an 
auxiliary bosonic field $\phi_{i}$. Choosing 
$b_i\equiv i n_i$, $A_{ij}\equiv v^{nl}_{ij}$ and $x_i\equiv \phi_i$, the 
transformation (\ref{eq:HS}) applied to the action (\ref{eq:U-V_action}) 
leads to
\begin{align}
&S[c^{*},c,\phi] = \int_{0}^{\beta}d\tau\Bigg\{ -\sum_{ij\sigma}c_{i\sigma}^{*}(\tau)\left[G_{0}^{-1}\right]_{ij}c_{j\sigma}(\tau)\nonumber\\
&\hspace{33mm}+
\alpha U\sum_{i}n_{i\uparrow}(\tau)n_{i\downarrow}(\tau)\Bigg\} \nonumber \\
&+ \int_{0}^{\beta}d\tau\Bigg\{ \frac{1}{2}\sum_{ij}\phi_{i}(\tau)[(v^{nl})^{-1}]_{ij}\phi_{j}(\tau)+i \alpha
\sum_{i}\phi_{i}(\tau)n_{i}(\tau)\Bigg\},
\label{eq:U-V_action_decoupled}
\end{align}
where we introduced the non-interacting fermionic lattice Green 
function $G_0$ defined by 
$\left[G_{0}^{-1}\right]_{ij}\equiv((-\partial_{\tau}+\mu)\delta_{ij}-t_{ij})$.
Again, a coupling constant $\alpha$ was introduced, and the physical
case corresponds to $\alpha$=1. Now, however, the coupling constant 
is not only a switch for turning on or off the fermion-boson coupling
but at the same time also the local Hubbard interaction.

In principle, the interaction should be a positive definite matrix in order for the Gaussian integrals invoked in the 
HS transformation to converge. Unlike the situation in the HS-UV decoupling where $U$ and $V$ are 
matrix elements of the screened Coulomb interaction, which is positive 
definite, this is not the case for the interaction of HS-V, $v^{nl}_{ij}$. 
This issue can be dealt with by adding an auxiliary identity matrix 
multiplied by a large enough constant.\cite{Sun_2002}  In practice, however, the simulation results are not 
affected by the value of this constant.

As before, the generating functional of correlation functions is 
obtained by introducing source terms. 
The fermionic Green's functions for this action is
unchanged compared to the UV-decoupling case:
$G_{ij}(\tau-\tau')=-\langle Tc_{i}(\tau)c_{j}^{\dagger}(\tau')\rangle=\delta\Omega/\delta J_{\mathrm{f},ij}(\tau,\tau')$.
The bosonic propagator formally still reads
$D_{ij}(\tau-\tau')=\langle T\phi_{i}(\tau)\phi_{j}(\tau')\rangle
=-2\delta\Omega/\delta J_{\mathrm{b},ij}(\tau,\tau')$.
It does not, however, correspond to the screened interaction, as in the HS-UV scheme:
in the case of vanishing fermion-boson coupling, the
bosonic propagator reduces by construction to only the non-local
part of the bare interaction.

The construction of the free energy functional $\Gamma$ proceeds
as before by Legendre transformation with respect to the sources 
$J_{\mathrm{f}}$ and $J_{\mathrm{b}}$, 
\begin{equation}
\Gamma_V[G,D]=\Omega[J_{\mathrm{f}}[G],J_{\mathrm{b}}[D]]-\mathrm{Tr}J_{\mathrm{f}}G
+\frac{1}{2}\mathrm{Tr}J_{\mathrm{b}}D,
\end{equation}
 with the reciprocity relations $J_{\mathrm{f}}=-\frac{\delta\Gamma_V}{\delta G}$
and $J_{\mathrm{b}}=2\frac{\delta\Gamma_V}{\delta D}$. 
The physical Green's functions will be obtained by setting $J_{\mathrm{f}}=0$ 
and $J_{\mathrm{b}}=0$, or equivalently,
by requiring the stationarity of $\Gamma_V$ with respect to $G$ and
$D$. 
Thanks to the choice of the coupling constant $\alpha$ in
front of the interaction and boson-fermion coupling terms,
$\alpha\left(U\sum n_{i\uparrow}n_{i\downarrow}+i\sum_{i}\phi_{i}n_{i}\right)$,
$\Gamma_V$ acquires the same form as before,
$\Gamma_{V,\alpha=1}=\Gamma_{V,\alpha=0}+\Psi_V$,
with $\Psi_V\equiv\int_{0}^{1}d\alpha\frac{d\Gamma_V}{d\alpha}$,
but it is now a functional of $G$ and $D$.

The non-interacting ($\alpha=0$) part of the $\Gamma$-functional
reads
\begin{eqnarray}
&&\Gamma_{V,\alpha=0}=\mathrm{Tr}\ln (-G)-\mathrm{Tr}(G_{0}^{-1}-G^{-1})G \nonumber\\
&&\hspace{12mm}-\frac{1}{2}\mathrm{Tr}\ln W+\frac{1}{2}\mathrm{Tr}(
(v^{nl})^{-1}-D^{-1})D.\label{eq:BK_alpha_0}
\end{eqnarray}

Finally, stationarity of the full $\Gamma_V$ reproduces the
fermionic Dyson equation for the Green's function and self-energy.
For the bosonic part, however, we obtain
$0=-\frac{1}{2}(D^{-1}-(v^{nl})^{-1})+\frac{\delta \Psi_V} { \delta D}$ for $D$. 
The bosonic self-energy
\begin{eqnarray}
\Pi_V =  -2\frac{\delta\Psi_V}{\delta D},
\end{eqnarray}
is thus not equal to the physical polarization of the system.

Again, solving the lattice problem defined by Eq.~(\ref{eq:U-V_action_decoupled}) amounts to evaluating the 
corresponding $\Psi_V$-functional, from which $\Sigma$ and $\Pi_V$,  and 
in turn $G$ and $D$ can be derived. 
Compared to the previous case of the $\Psi$-functional, however,
the sublety of $D$ being the screened non-local interaction
(not equal to the full $W$) requires additional care in the construction of a combined DMFT scheme.

\section{Methods of Solution}
\label{Methods-of-Solution}
\subsection{EDMFT}

$\Psi$ is a functional of the fermionic and bosonic Green functions $G_{ij}$ and $W_{ij}$. EDMFT approximates $\Psi\left[G_{ij},W_{ij}\right]$ by $\Psi^{EDMFT}\left[G_{ii},W_{ii}\right]=\Psi\left[G_{ii}\delta_{ij},W_{ii}\delta_{ij}\right]$. 
Similarly, $\Psi_V^{EDMFT}\left[G_{ii},D_{ii}\right]=\Psi_V\left[G_{ii}\delta_{ij},D_{ii}\delta_{ij}\right]$. 
The local Green functions can be obtained by solving an auxiliary effective local problem defined by the action
\begin{widetext}
\begin{eqnarray}
 S_\text{eff,HS-UV}^{EDMFT} & = & -\int_{0}^{\beta}d\tau d\tau'\sum_{\sigma}c_{\sigma}^{\dagger}(\tau)\mathcal{G}^{-1}(\tau-\tau')c_{\sigma}(\tau')\nonumber\\
 & + & \frac{1}{2}\int_{0}^{\beta}d\tau d\tau'\phi(\tau)\mathcal{U}^{-1}(\tau-\tau')\phi(\tau')+i\int_{0}^{\beta}d\tau\phi(\tau)n(\tau)\label{eq:Effective_action_local_HS-UV} \\
S_\text{eff,HS-V}^{EDMFT} & = & -\int_{0}^{\beta}d\tau d\tau'\sum_{\sigma}c_{\sigma}^{\dagger}(\tau)\mathcal{G}^{-1}(\tau-\tau')c_{\sigma}(\tau')+\int_{0}^{\beta}d\tau Un_{\uparrow}(\tau)n_{\downarrow}(\tau)\nonumber \\
 & + & \frac{1}{2}\int_{0}^{\beta}d\tau d\tau'\phi(\tau)\mathcal{D}^{-1}(\tau-\tau')\phi(\tau')+i\int_{0}^{\beta}d\tau\phi(\tau)n(\tau).\label{eq:Effective_action_local_HS-V}
\end{eqnarray}

\end{widetext}
This action is very similar to the lattice action (Eq.~(\ref{eq:U-V_action_decoupled})), with $G_{0}$ replaced by an appropriately defined dynamical $\mathcal{G}$ decribing the
excursions of an electron in the lattice from a given site (the impurity, defined by operators $c_{\sigma}$ and $c_{\sigma}^{\dagger})$ and back,
and the bare and instantaneous interaction $v$ (or $ v^{nl}$ in HS-V) replaced by the retarded interaction
$\mathcal{U}$ (or $\mathcal{D}$). The effective actions (\ref{eq:Effective_action_local_HS-UV}) and (\ref{eq:Effective_action_local_HS-V}) are obtained by integrating out all sites but one in the lattice action and taking the infinite-dimensional limit. The derivation of the action for the HS-UV scheme, as well as the EDMFT loop sketched below, are presented in Appendix \ref{cavity_method}.

Integrating out the $\phi$-field in Eqs.~(\ref{eq:Effective_action_local_HS-UV}) and (\ref{eq:Effective_action_local_HS-V})
yields the impurity actions
\begin{widetext}
\begin{align}
S_\text{eff, HS-UV}^{EDMFT} &=  -\int_{0}^{\beta}d\tau d\tau' \sum_{\sigma}c_{\sigma}^{\dagger}(\tau)\mathcal{G}^{-1}(\tau-\tau')c_{\sigma}(\tau')
+\frac{1}{2}\int_{0}^{\beta}d\tau d\tau' n(\tau)\mathcal{U}(\tau-\tau')n(\tau')-\frac{1}{2}\mathrm{Tr}\ln\mathcal{U},\label{eq:Effective_action_1} \\
S_\text{eff, HS-V}^{EDMFT} &=  -\int_{0}^{\beta}d\tau d\tau'\sum_{\sigma}c_{\sigma}^{\dagger}(\tau)\mathcal{G}^{-1}(\tau-\tau')c_{\sigma}(\tau')
+\int_0^{\beta} d\tau Un_{\uparrow}(\tau)n_{\downarrow}(\tau)
+\frac{1}{2}\int_{0}^{\beta}d\tau d\tau' n(\tau)\mathcal{D}(\tau-\tau')n(\tau')-\frac{1}{2}\mathrm{Tr}\ln\mathcal{D},\label{eq:Effective_action_2}
\end{align}
\end{widetext}
which feature a retarded retarded interaction $\mathcal{U}(\tau-\tau')$ (for HS-UV) or $\mathcal{D}(\tau-\tau')$ (for HS-V) between charges.

The solution of this impurity problem, described in detail in Section~\ref{numerical}, requires the calculation of the one-particle Green's functions $G_\text{loc}\equiv-\langle T c(\tau)c^{\dagger}(0)\rangle$ and $W_\text{loc}\equiv\langle T \phi(\tau)\phi(0)\rangle$ (or $D_\text{loc}$ for HS-V). 
From $G_\text{loc}$ and $W_\text{loc}$, one computes the corresponding self-energies, $\Sigma_\text{loc} = \mathcal{G}^{-1}-G_\text{loc}^{-1}$ and $\Pi_\text{loc} = \mathcal{U}^{-1}-W_\text{loc}^{-1}$  (or $(\Pi_V)_\text{loc} = \mathcal{D}^{-1}-D_\text{loc}^{-1}$).
The EDMFT approximation consists in identifying the impurity self-energies with the lattice self-energies: $\Sigma(k,i\omega_n)\approx \Sigma_\text{loc}(i\omega_n)$, $\Pi(k,i\nu_n) \approx \Pi_\text{loc}(i\nu_n)$. This allows one to evaluate (approximate) lattice Green functions $G(k,i\omega_n)$ and $W(k,i\nu_n)$ (or $D(k,i\nu_n)$) through Dyson's equation and estimates of the local lattice Green functions by summation over $k$. Eventually, one obtains updated $\mathcal{G}$ and $\mathcal{U}$ (or $\mathcal{D}$) via

\begin{eqnarray} 
\mathcal{G}^{-1} & = & G_\text{loc}^{-1}[\Sigma_\text{loc}]+\Sigma_\text{loc},\label{eq:Self_consistency_G}\\
\mathcal{U}^{-1} & = & W_\text{loc}^{-1}[\Pi_\text{loc}]+\Pi_\text{loc}\label{eq:Self_consistency_W},\\
\mathcal{D}^{-1} & = & D_\text{loc}^{-1}[(\Pi_V)_\text{loc}]+(\Pi_V)_\text{loc}\label{eq:Self_consistency_D}.
\label{eq:Self_consistency_U} 
\end{eqnarray}

\subsection{The GW approximation\label{GW_approximation}}

While EDMFT can treat strong local correlations, it completely neglects the nonlocal contributions to the self-energy. A complementary approach, which treats spatial fluctuations, but works reliably only in the weakly correlated regime is the GW method.\cite{Hedin_1965}  
The GW approximation has been used extensively to investigate the properties of weakly correlated materials,
such as the band gaps of semi-conductors (see \emph{e.g.} Ref.~\onlinecite{Shishkin_2007}).
In these materials GW correctly accounts for the screening effects
of the electrons at the random-phase approximation (RPA) level. Schematically,
for a general Coulomb interaction $v(r)\sim1/r$, one-shot GW consists in replacing
the bare interaction $v$ of the Fock self-energy $\Sigma_{F}\sim G_0 v$
by the screened interaction $W=v/\epsilon_{RPA}$, where $\epsilon_{RPA}=1-vP_{0}$
and $P_{0}\sim G_{0}G_{0}$ is the dynamical Lindhard function. In a self-consistent scheme, $G_0$ is replaced by the interacting Green function $G$. The Fock self-energy
thus becomes $\Sigma_{GW}\sim GW$, hence the name of the approximation.
Formally, the GW approximation can be obtained by Hubbard-Stratonovich
decoupling the Coulomb interaction $v$ \emph{via} an auxiliary bosonic
field $\phi$ characterized by the propagator $W\sim\langle\phi\phi\rangle$.
This amounts to replacing the electron-electron interaction by the indirect interaction of two electrons mediated by a boson described by $\phi$. The first-order self-energy
diagram in the expansion of this electron-boson interaction is $\Sigma\sim GW$.

\begin{figure}[h]
\begin{centering}
\includegraphics[width=0.08 \textwidth]{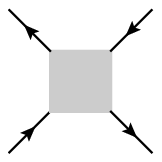}
\hfill{}
\includegraphics{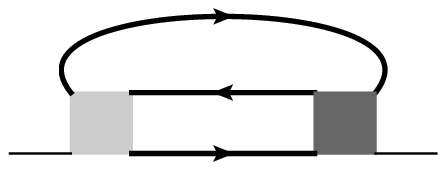}
\includegraphics[width=0.45 \textwidth]{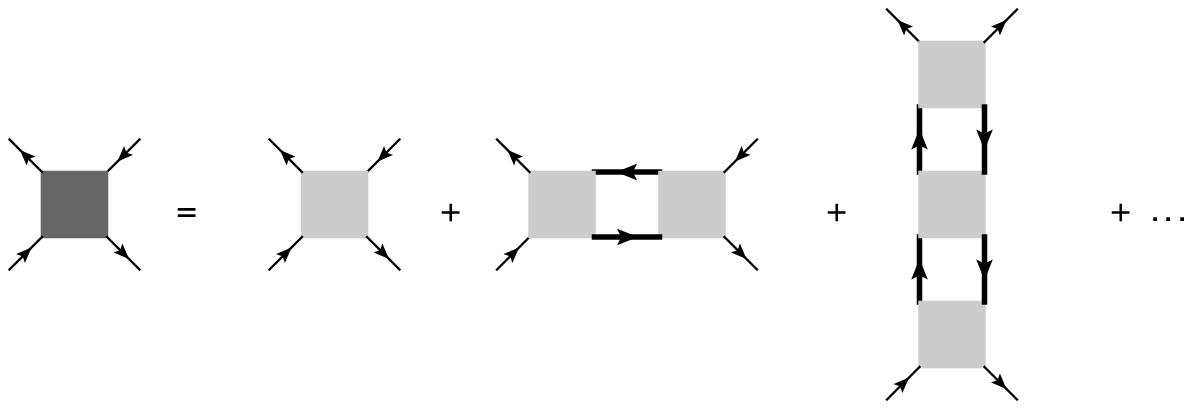}
\end{centering}
\caption{\label{fig:ee_expansion}
Diagrammatic expansion of the electron-electron vertex. \emph{From left to right, top to bottom}:
Bare electron-electron interactionvertex $\Gamma_{\mathrm{ee}}^{(0)}$.
Fully boldified $\Sigma_{\mathrm{ee}}$.
Expansion of the full electron-electron vertex, $\Gamma_{\mathrm{ee}}$.
}
\end{figure}

\begin{figure}[h]
\begin{centering}
\includegraphics[width=0.1\textwidth]{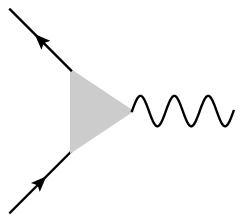}\\
\includegraphics[width=0.2\textwidth]{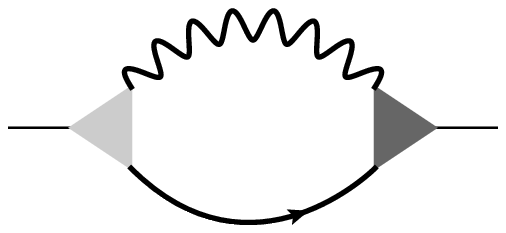}
\hfill{}
\includegraphics[width=0.2\textwidth]{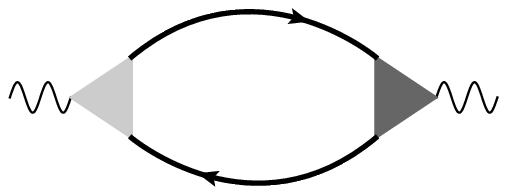}
\includegraphics[width=0.45\textwidth]{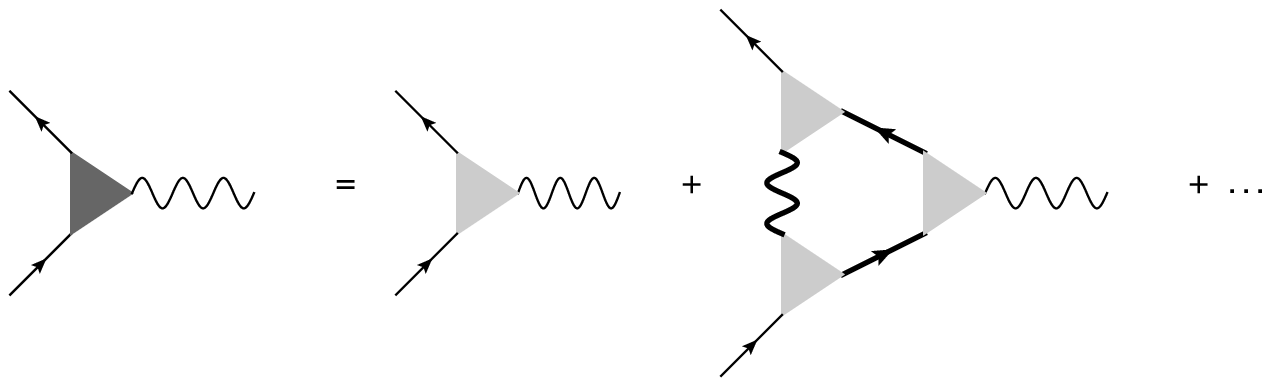}
\end{centering}
\caption{\label{fig:ep_expansion}
Diagrammatic expansion for an electron-boson vertex. \emph{From left to right, top to bottom}:
Bare electron-boson interaction vertex $\Gamma_{\mathrm{eb}}^{(0)}$. 
Fully boldified $\Sigma_{\mathrm{eb}}$. 
Fully boldified $\Pi$. 
Expansion of the full electron-boson vertex, $\Gamma_{\mathrm{eb}}$. 
}
\end{figure}

The standard derivation of the GW approximation relies on a 
truncation of Hedin's equations,\cite{Hedin_1965} where the three-legged vertex
$\Lambda = 1 + \frac{\delta \Sigma}{\delta G}GG\Lambda$ is set to unity.
In the following, we will derive the GW approximation for our lattice
model in a diagrammatic way based on the four-legged vertices of
standard perturbation theory, for both free energy functionals,
that is, both choices of the HS decoupling. 
In the HS-V approach, the lattice action of Eq.~(\ref{eq:U-V_action_decoupled})
contains two interaction vertices: a local electron-electron interaction
$Un_{i\uparrow}n_{i\downarrow}$ and a local electron-boson interaction
$i\phi_{i}n_{i}$. Consequently, the perturbation expansion of the
Green's function will contain two types of bare interaction vertices,
namely $\Gamma_{\mathrm{ee}}^{(0)}(\tau_{1},\tau_{2},\tau_{3},\tau_{4})_{ijkl}=U\delta_{ijkl}\delta_{i\uparrow}\delta_{j\downarrow}\delta_{k\uparrow}\delta_{l\downarrow}\delta(\tau_{1}-\tau_{2})\delta(\tau_{3}-\tau_{4})\delta(\tau_{2}-\tau_{3})$
and $\Gamma_{\mathrm{eb}}^{(0)}(\tau_{1},\tau_{2},\tau_{3})_{ijk}=i\delta_{ijk}\delta(\tau_{1}-\tau_{2})\delta(\tau_{3}-\tau_{2})$,
which we will represent as shown in Figs.~\ref{fig:ee_expansion}
and \ref{fig:ep_expansion}. We will suppose that we can perform
the expansions separately and then sum the two results (which is an approximation since there could well be sequences of interactions
with alternating $\Gamma_{\mathrm{ee}}^{(0)}$ and $\Gamma_{\mathrm{eb}}^{(0)}$). In the HS-UV approach, there is only the electron-boson vertex $\Gamma_{\mathrm{eb}}^{(0)}$.

The vertex $\Gamma_{\mathrm{ee}}^{(0)}$ will lead to contributions that are not present in the HS-UV decoupling scheme.  The perturbation expansion in powers of $\Gamma_{\mathrm{ee}}^{(0)}$ yields a series of self-energy diagrams, whose lines are non-interacting Green's functions $G_0$. 
Since some higher-order diagrams contain ``self-energy insertions", the number of diagrams can
be reduced by {}``boldifying'' the lines, namely by replacing $G_{0}$ by the interacting Green's function $G$.
Subsequently, the number of diagrams can be further reduced by regrouping the interaction vertices into
a {}``boldified'' vertex $\Gamma_{\mathrm{ee}}$ pictured in Fig.~\ref{fig:ee_expansion}.
Thus the electron-electron part of the self-energy (beyond the Hartree self-energy) can be described (Fig.~\ref{fig:ee_expansion}) by the exact expression\cite{Abrikosov_1965}
\begin{eqnarray}
\Sigma_{\mathrm{ee}}=\Gamma_{\mathrm{ee}}^{(0)}GGG\Gamma_{\mathrm{ee}},
\end{eqnarray}
with bold propagators $G$ \emph{and} bold vertices $\Gamma_{\mathrm{ee}}$.  Note
that only the right vertex is boldified to avoid double-counting. This vertex is also called the fully reducible vertex. We stress that this electron-electron contribution to the electronic self-energy is absent in the HS-UV approach, since in this scheme there is no longer any electron-electron interaction vertex after the HS decoupling.

Similarly, the expansion of the partition function in powers of $\Gamma_{\mathrm{eb}}^{(0)}$ can be simplified by boldifying the $G_{0}$ and $W_{0}$ ($=v$ or $v^{nl}$) lines and the vertices, leading to graphs of the form displayed in Fig.~\ref{fig:ep_expansion}, corresponding to the expressions
\begin{eqnarray}
\Sigma_{\mathrm{eb}} & = & \Gamma_{\mathrm{eb}}^{(0)}GW\Gamma_{\mathrm{eb}},\\
\Pi_{\mathrm{eb}} & = & -2 \Gamma_{\mathrm{eb}}^{(0)}GG\Gamma_{\mathrm{eb}}.
\end{eqnarray}
The expansion for the bold electron-boson vertex $\Gamma_{\mathrm{eb}}$ is also pictured in Fig.~\ref{fig:ep_expansion}.

The usual GW approximation consists in truncating the expansion of the electron-boson vertex
function $\Gamma_{\mathrm{eb}}$ after its first term,
namely by taking $\Gamma_{\mathrm{eb}}\approx\Gamma_{\mathrm{eb}}^{(0)}=i\delta$. This simplification, which amounts to neglecting the so-called ``vertex corrections", yields the familiar expressions $\Sigma_{\mathrm{eb}}^{GW}=-GW$ and $\Pi^{GW}=2 GG$. For the HS-UV decoupling, $\Sigma = \Sigma_{\mathrm{eb}}$ and thus 
\begin{eqnarray}
\Sigma^{GW}_{HS-UV} = -G W,
\label{eq:GW_HS-UV}
\end{eqnarray}
For the HS-V decoupling,  there is a second contribution coming from the electron-electron vertex. If one approximates $\Gamma_{\mathrm{ee}}\approx \Gamma_{\mathrm{ee}}^{(0)}$ (as in Ref.~\onlinecite{Sun_2002}), one gets $\Sigma_{\mathrm{ee}}^{GW} =- U^{2}GGG$, and hence
\begin{eqnarray}
\Sigma^{GW}_{HS-V} = -GD-U^{2}GGG.
\label{eq:GW_HS-V}
\end{eqnarray}

At this point, a few remarks are in order: The two approximations (\ref{eq:GW_HS-UV}) and (\ref{eq:GW_HS-V}) are not equivalent.
Making the lowest-order approximation on the electron-electron vertex is a stronger assumption than truncating the electron-boson vertex. In the HS-V approach, the series of diagrams corresponding to Eq.~(\ref{eq:GW_HS-V}) contains the ring of ``bubbles" made up of $G$ and $v^{nl}_{ij}$ (which contains only the off-site repulsion $V$), plus a second order diagram in $U$. In contrast, the diagrams in $\Sigma^{GW}_{HS-UV}$ contain the ring of bubbles made up of $G$ and $v_{ij}$ (which contains $U$ and $V$). Put differently, it not only comprises the off-site interaction $V$ to all orders (at the RPA level), but also the on-site interaction $U$ to all orders (at the RPA level). We thus expect the HS-UV scheme to be better poised to capture non-local effects arising from $V$ \emph{and} $U$, while the HS-V scheme will probably give non-trivial contributions only in parameter regimes where $V$ plays the dominant role.
While we benchmark both approaches in the results section,
we therefore focus in the following discussion of the
combined GW+DMFT scheme on the formulation in terms of
the Almbladh functional. A similar combination based
on the $\Psi_V$ functional is possible, leading to
a combination of the $GD$+self-consistent 2nd order perturbation theory
expression of Eq. (\ref{eq:GW_HS-V}) with dynamical
mean field theory, as described in Ref.~\onlinecite{Sun_2002}.
We refer to this combination in the following as
``GD+SOPT+DMFT''.

\subsection{GW+DMFT approach}

As already hinted at before, EDMFT and GW are complementary approximate schemes: EDMFT provides a good description of local correlations, while GW captures longer-range correlations and in particular long-range screening. Therefore, combining both approximations appears promising. The GW+DMFT approach \cite{Aryasetiawan_Biermann_2004,Biermann_2003} consists in making an approximation on $\Psi[G_{ij},W_{ij}]$ by decomposing it in the following way: 
\begin{eqnarray}
 \Psi\approx\Psi^{EDMFT}[G_{ii},W_{ii}]+\Psi^{GW}_{\mathrm{nonloc}}[G_{ij},W_{ij}]
 \end{eqnarray}
where $\Psi^{GW}_{\mathrm{nonloc}} = \Psi^{GW} - \Psi^{GW}_{\mathrm{loc}}$.

While EDMFT will generate the series of local self-energy diagrams up to infinite order, the non-local contributions to $\Psi$ will be generated in a perturbative way by the non-local part of GW diagrams, thus avoiding double-counting.

In the limit of infinite dimensions, non-local diagrams vanish. Thus, the effect (if any) of the non-local contributions is expected to occur only as the dimension is lowered. The proximity to a phase transition should also enhance spatial fluctuations. In principle, the GW contribution should nonetheless remain a correction
to the DMFT part, which justifies why $\Psi_{\mathrm{nonloc}}[G_{ij},W_{ij}]$ can be treated on a perturbative level, while $\Psi[G_{ii},W_{ii}]$ is evaluated exactly. 

The approximate electronic self-energy will be given by $\Sigma_{ij}=\Sigma_{ii}+(1-\delta_{ij})\Sigma_{ij}$
where $\Sigma_{ii}=\Sigma_\text{loc}$ and $\Sigma_{ij}=\Sigma_{ij}^{GW}$. The $1-\delta_{ij}$ factor
ensures the substraction of the local part of the GW self-energy. Analogous expressions hold for $\Pi_{ij}$.

This approach is very general. In the specific case of the extended Hubbard model, one can expect the GW contribution to be significant as one approaches an instability in the charge sector, namely close to the charge-ordering transition. The GW diagrams can in principle be replaced by any other perturbative diagrammatic correction, corresponding to decoupling the interaction in other channels.

\section{Numerical Implementation}
\label{numerical}

\subsection{Solution of the EDMFT impurity problem}

\begin{figure}[h]
\begin{centering}
\includegraphics[width=0.45\textwidth]{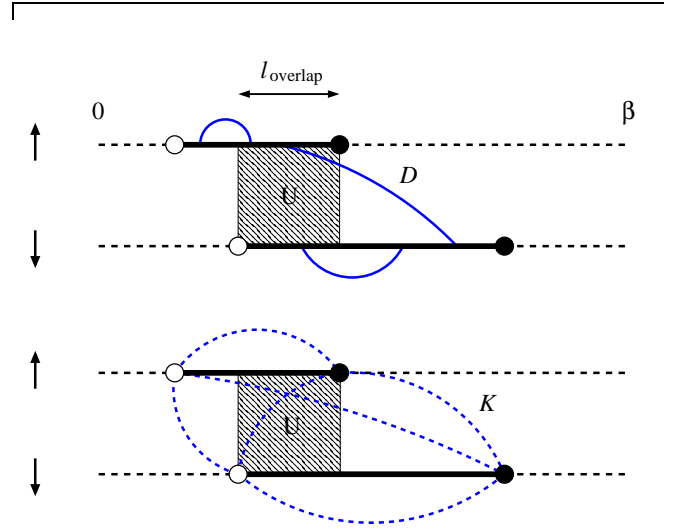}
\end{centering}
\caption{\label{fig:segments} (Color online). 
Illustration of a Monte Carlo configuration in the segment representation. The top figure corresponds to a configuration with one spin-up and one spin-down segment, each representing a time-interval marked by a creation operator (empty circle) and an annihilation operator (full circle). The overlap of the segments, corresponding to the width of the hashed region, yields the weight due to the instantaneous interaction $U$. The retarded interaction $\mathcal{D}(\tau-\tau')$ is represented by the blue curved lines.   
The bottom panel represents the weight of this configuration after integration of the retarded interaction over the segments. The dashed blue lines correspond to the interaction $K(\tilde \tau_i-\tilde \tau_j)$ between creation and annihilation operators.
}
\end{figure}

The impurity models~(\ref{eq:Effective_action_1}) and (\ref{eq:Effective_action_2}) can be solved efficiently using the hybridization-expansion
continuous-time quantum Monte-Carlo solver (CTQMC-hyb).\cite{Werner_2006}
The formalism has been previously derived using a Hamiltonian representation of the impurity model.\cite{Werner_2007,Werner_2010}
Here, we discuss an alternative derivation based on the effective action, focussing on the case of action~(\ref{eq:Effective_action_1}).
A CTQMC-hyb simulation consists of sampling configurations representing
specific time-sequences of ``hybridization events",  with weight proportional to the determinant of a matrix  of
hybridization functions.
The perturbation expansion of the partition function $Z$ and the summation of diagrams with identical operator sequences leads to
\begin{widetext}
\begin{eqnarray}
Z & = & \sum_{\{n_{\sigma}\}=0}^{\infty}\prod_{\sigma}\left[\frac{1}{(n_\sigma!)^2}\int_{0}^{\beta}d\tau^\sigma_{1}\int_{0}^{\beta}d{\tau'}^\sigma_{1}\dots\int_{0}^{\beta}d\tau^\sigma_{n_{\sigma}}\int_{0}^{\beta}d{\tau'}^\sigma_{n_{\sigma}}\mathrm{Det}\Delta_{\sigma}\right]\nonumber\\
 &  & \times \int D[c^*,c] e^{-S_{at}}T \prod_\sigma c_{\sigma}^{*}({\tau'}^\sigma_{1})c_{\sigma}(\tau^\sigma_{1})\dots c_{\sigma}^{*}({\tau'}^\sigma_{n_{\sigma}})c_{\sigma}(\tau^\sigma_{n_{\sigma}}),
 \label{eq:hyb_exp}
\end{eqnarray}
\end{widetext}
where $(\Delta_{\sigma})_{ij}=\Delta_{\sigma}(\tau_{i}-\tau_{j}')$ 
is the hybridization function evaluated for the time-difference between annihilation operator $i$ and creation operator $j$
and

\begin{eqnarray}
S_{at}&=&\frac{1}{2} \sum_{\sigma \sigma'} \int\int d\tau d\tau'  n_{\sigma}(\tau)\mathcal{U}_{\sigma\sigma'}(\tau-\tau') n_{\sigma'}(\tau')\nonumber\\
&+& \int d\tau\sum_{\sigma}c_{\sigma}(\tau)^{*}\left(\partial_{\tau}-\mu\right)c_{\sigma}(\tau)
\end{eqnarray}

represents the interaction and chemical potential contributions of action (\ref{eq:Effective_action_2}).
The interaction $\mathcal{U}$ can always be split into a delta-function contribution and a non-singular contribution, $\mathcal{U(\tau)}_{\sigma,\sigma'} = U\delta(\tau)(1-\delta_{\sigma,\sigma'})+\mathcal{D}(\tau)_{\sigma,\sigma'}$. (In the HS-V scheme, this separation is already explicit.)
The last factor of Eq.~(\ref{eq:hyb_exp}) can be easily evaluated since the time evolution operators are diagonal in the occupation number basis. In the segment representation\cite{Werner_2006} each imaginary time interval with occupation $n_\sigma=1$ is marked by a segment, and the last factor can (up to a permutation sign) be written as $w_{\mu}w_U w_{\mathcal{D}}$ with $w_{\mu} = e^{\mu(l_{\uparrow}+l_{\downarrow})}$ and $w_{U} = e^{-U l_{\mathrm{overlap}}}$. Here, $l_{\sigma}$ stands for the total length of segments of spin $\sigma$, while $l_{\mathrm{overlap}}$ is the total overlap between segment of opposite spin (see illustration of a segment configuration in Fig.~\ref{fig:segments}). The retarded interaction contributes a factor
\begin{align}
& w_{\mathcal{D}}  = e^{-\frac{1}{2}\sum_{\sigma_{1},\sigma_{2}}\int_{0}^{\beta}d\tau_{1}\int_{0}^{\beta}d\tau_{2}\mathcal{D}(\tau_{1}-\tau_{2})_{\sigma_1,\sigma_2}n_{\sigma_{1}}(\tau_{1})n_{\sigma_{2}}(\tau_{2})}\nonumber
\\
 & =  \exp\left(-\frac{1}{2}\sum_{\substack{\sigma_{1}\sigma_{2}\\k_{\sigma_{1}}k_{\sigma_{2}}}}\int_{\tau_{k_{\sigma_{1}}}}^{\tau_{k_{\sigma_{1}}}'}d\tau_{1}\int_{\tau_{k_{\sigma_{2}}}}^{\tau_{k_{\sigma_{2}}}'}d\tau_{2}\mathcal{D}(\tau_{1}-\tau_{2})_{\sigma_1,\sigma_2}\right)\nonumber \\
 \label{u_w}
\end{align}
where $\{k_{\sigma}\}$ represents the collection of segments of spin $\sigma$.

Let us now define a function $K(\tau)$ such that $K''(\tau) = \mathcal{D}(\tau)$ for $0<\tau<\beta$ and $K(0)=K(\beta)=0$. $K$ is $\beta$-periodic and symmetric around $\tau=\beta/2$. It has a slope discontinuity at zero, so that the second derivative also gives a delta-function contribution. Carrying out the integral in Eq.~(\ref{u_w}) thus yields
\begin{align}
\ln w_{\mathcal{D}}=&-\frac{1}{2}\sum_{\substack{\sigma_{1}\sigma_{2}\\k_{\sigma_{1}}k_{\sigma_{2}}}} \Big[-K(\tau_{k_{\sigma_{1}}}'-\tau_{k_{\sigma_{2}}}')+K(\tau_{k_{\sigma_{1}}}-\tau_{k_{\sigma_{2}}}')\nonumber\\
&\hspace{1.9cm}+K(\tau_{k_{\sigma_{1}}}'-\tau_{k_{\sigma_{2}}})-K(\tau_{k_{\sigma_{1}}}-\tau_{k_{\sigma_{2}}})\Big]\nonumber\\
&+K'(0)(l_\uparrow+l_\downarrow)+2K'(0)l_{\text{overlap}}\phantom{\Bigg]}.
\end{align}
Using the fact that $K(\tau)$ is an even function, we can write $\ln{w_{\mathcal{D}}}=\sum_{i>j}s_i s_j [K(\tilde \tau_i-\tilde \tau_j)-K(0)] +K'(0)(l_\uparrow+l_\downarrow)+2K'(0)l_{\text{overlap}}$ where the time-arguments of the hybridization events (creation and annihilation operators) are now ordered as $0<\tilde \tau_1< \tilde \tau_2 < \ldots <\beta$ and $s$ is $+1$ for a creation operator and $-1$ for an annihilation operator. 

We conclude that the retarded part of  the interaction $\mathcal{D}(\tau-\tau')$ results in a retarded ``interaction'' between all pairs of impurity creation and annihilation operators, as well as a shift of the instantaneous interaction $U\rightarrow \tilde U=U-2K'(0)$ and a shift of the chemical potential $\mu\rightarrow \tilde \mu = \mu + K'(0)$. If one writes the interaction term in terms of density fluctuations, $\frac{1}{2}\int\int \bar{n}(\tau)\mathcal{D}(\tau-\tau')\bar{n}(\tau')$ with $\bar{n}=n-\langle{n}\rangle$, the only change induced in the weight is yet another shift of the chemical potential $\tilde{\mu}=\mu +K'(0)-2\langle n\rangle K'(0)$. The retarded interactions can be evaluated at negligible computational cost since the calculation of this contribution for a local update is $O(n)$ (where $n$ is the number of operators), while the evaluation of a determinant ratio is $O(n^2)$.

In practice, the local bosonic propagator $W_\text{loc}=\langle T\phi(\tau)\phi(0)\rangle$ needed
in Eq.~(\ref{eq:Self_consistency_W}) is deduced from the connected charge-charge
correlation function $\chi_\text{loc}=\langle T \bar n(\tau) \bar n(0)\rangle$ via
the relation 

\begin{eqnarray}
W_\text{loc}=\mathcal{U}-\mathcal{U}\chi_\text{loc}\mathcal{U}.\label{eq:W_Chi_relation}
\end{eqnarray}
Indeed, using Eq.~(\ref{eq:Effective_action_local_HS-V}), $W_\text{loc}$ can
be reexpressed as $W_\text{loc}=-2\frac{\delta\ln Z}{\delta\mathcal{U}^{-1}}$.
The chain rule $\frac{\delta\ln Z}{\delta\mathcal{U}^{-1}}=-\mathcal{U}^{2}\frac{\delta\ln Z}{\delta\mathcal{U}}$
and Eq.~(\ref{eq:Effective_action_2}) give $\frac{\delta\ln Z}{\delta\mathcal{U}}=-\frac{1}{2}\chi_\text{loc}+\frac{1}{2}\mathcal{U}^{-1}$, and hence one arrives at Eq.~(\ref{eq:W_Chi_relation}). An analogous expression holds for $D_\text{loc}$.

\subsection{Self-consistency}

The EDMFT+GW scheme is generally expected to work well if the non-local GW contributions to the self-energy is a relatively small correction to the local self-energy computed by EDMFT. It thus makes sense to first obtain a reasonable guess of the final solution by applying EDMFT only, and then compute the non-local correction and study its effect on the properties of the system. Following this observation, we implemented the EDMFT+GW scheme as follows: for a given $U$ and $V$, we
(i) obtain a converged EDMFT solution,
(ii) take the EDMFT solution as the starting point for a self-consistent EDMFT+GW calculation, and
(iii) stop when local and non-local observables have converged.

This is not the only possible combination of EDMFT with GW, although the final result of the self-consistent scheme should not depend on the starting point, provided the scheme converges. For instance, one can choose to initiate the scheme by computing $\Sigma^{GW}$ and $\Pi^{GW}$ from non-interacting propagators $G_{0}(k,i\omega_n)=[i\omega_n+\mu-\epsilon(k)]^{-1}$ and $W_0(k,i\nu_n)=v_k$. Yet, these propagators yield a very large GW polarization owing to their metallic character (they correspond to the $U=0$, $V=0$ case). Especially for the regions of interest here ($V$ close to $V_c$ and finite $U$), this large polarization is far from the expected solution. Indeed, one can observe that when taking an insulating $G$ and \emph{e.g.} $W=v$ as inputs for GW, the resulting polarization is small compared to the local polarization $\Pi_\text{loc}$.

\subsection{Analytical continuation\label{analytical_continuation}}

The non-trivial structures of the frequency-dependent interaction
result in additional features in the local spectral function $A(\omega)=-\frac{1}{\pi}\mathrm{Im}G(\omega+i\eta)$.
For example, the
case $\mathrm{Im}\mathcal{D}(\omega)=-\lambda^2\left(\delta(\omega-\omega_{0})-\delta(\omega+\omega_{0})\right)$,
studied in Ref.~\onlinecite{Werner_2010}, corresponds to the Holstein-Hubbard
model, for which the local spectral function is expected to display
plasmonic peaks at multiples of the {}``plasmon'' frequency $\omega_{0}$.\cite{Casula_2012}
However, the commonly-used MaxEnt algorithm \cite{Jarrell_1996}
tends to smooth out high-energy features and therefore a dedicated
scheme must be implemented to recover the sought-after
features. We proceed as proposed in Ref.~\onlinecite{Casula_2012}:
(a) From $\mathcal{U}(i\nu_{n})$, we compute the bosonic function
$B(\tau)=\exp\left(\sum_{n\neq0}\frac{\mathcal{U}(i\nu_{n})-\mathcal{U}(0)}{\nu_{n}^{2}}\left(e^{i\tau\nu_{n}}-1\right)\right)$,
its Fourier transform $B(i\nu_{n})$ and, using a Pad\'e procedure,\cite{Vidberg_1977}
its spectral function $B(\omega)$; (b) From $G(\tau)$, we compute an
auxiliary function $G_\text{aux}(\tau)=G(\tau)/B(\tau)$ and use MaxEnt
to obtain $A_\text{aux}(\omega)$; (c) Finally, we compute the spectral function as the convolution
\begin{equation}
A(\omega)=\int_{-\infty}^{\infty}d\epsilon B(\epsilon)\frac{1+e^{-\beta\omega}}{(1+e^{\beta(\epsilon-\omega)})(1-e^{-\beta\epsilon})}A_\text{aux}(\omega-\epsilon).
\label{eq:Convolution}
\end{equation}

\section{Summary of the Computational Scheme}

The computational scheme can be summarized as follows for the HS-UV (resp. HS-V) scheme:
\begin{enumerate}

\item Start with an initial guess for $\Sigma(k,i\omega)$ and $\Pi(k,i\nu)$:
for instance, $\Sigma=0$ and $\Pi=0$ (non-interacting limit).

\item \textbf{Lattice Green functions}: Compute $G(k,i\omega)$ and $W(k,i\nu) $ (resp. $D(k,i\nu) $) \emph{via} Dyson's equation with $v_k$ (resp. $v^{nl}_k$) as the bare interaction. 

\item \textbf{EDMFT self-consistency}: Extract $G_\text{loc}(i\omega)=\sum_{k}G(k,i\omega)$
and $W_\text{loc}(i\nu)=\sum_{k}W(k,i\nu)$ (resp. $D_\text{loc}$) and use Eqs. (\ref{eq:Self_consistency_W}) and  (\ref{eq:Self_consistency_G})
to find $\mathcal{G}(i\omega)$ and $\mathcal{U}(i\nu)$ (resp. $\mathcal{D}(i\nu)$).

\item \textbf{Impurity solver}: Compute $G_\text{loc}(\tau)$ and $\chi_\text{loc}(\tau)$ (resp. $\chi^V_\text{loc}(\tau)$),
as well as $W_\text{loc}=\mathcal{U}-\mathcal{U}\chi_\text{loc} \mathcal{U}$ (resp. $D_\text{loc}=\mathcal{D}-\mathcal{D}\chi^V_\text{loc} \mathcal{D}$). From these, extract the self-energies
$\Sigma_\text{loc}=\mathcal{G}^{-1}-G_\text{loc}^{-1}$ and
$\Pi_\text{loc}=\mathcal{U}^{-1}-W_\text{loc}^{-1}$ (resp. $(\Pi_V)_\text{loc}=\mathcal{D}^{-1}-D_\text{loc}^{-1}$).

\item \textbf{GW+DMFT step (optional):}

\begin{enumerate}

\item Compute\\ $\Pi^{GW}(k,\tau)=2\sum_{q}G(q,\tau)G(q-k,-\tau)$
and $\Sigma^{GW}(k,\tau)=-\sum_{q}G(q,\tau)W^c(k-q,\tau)$, where $W^c = W - v$ (resp. $\Sigma^{GW}(k,\tau)=-\sum_{q}G(q,\tau)D(k-q,\tau)-U^{2}\sum_{q}G(q,\tau)\Pi^{GW}(q-k,\tau)$).

\item Extract non-local parts from GW: $\Sigma_\mathrm{non\, local}^{GW}(k,i\omega)=\Sigma^{GW}(k,i\omega)-\sum_{k}\Sigma(k,i\omega)$
and $\Pi_\mathrm{non\, local}^{GW}(k,i\nu)=\Pi^{GW}(k,i\nu)-\sum_{k}\Pi(k,i\nu)$ 

\item Combine $\Sigma_\text{loc}(i\omega)$ and $\Sigma_\mathrm{non\, local}^{GW}(k,i\omega)$
into $\Sigma(k,i\omega)$, as well as $\Pi_\text{loc}(i\nu)$ and $\Pi_\mathrm{non\, local}^{GW}(k,i\nu)$
into $\Pi(k,i\nu)$. 

\end{enumerate}

\item Go back to step 2 until convergence.

\end{enumerate}

\begin{figure}[h]
\begin{centering}
\includegraphics[scale=0.43]{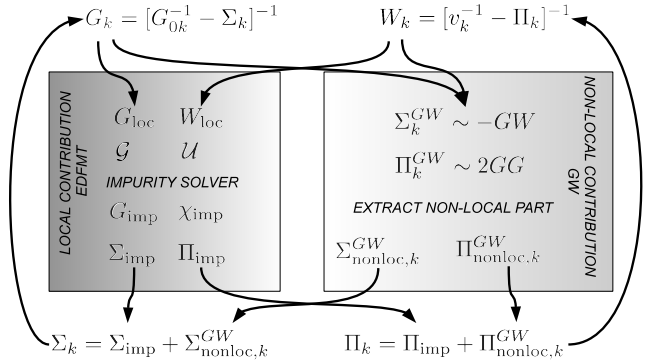}
\end{centering}
\caption{\label{fig:Computational-scheme}
Computational scheme (HS-UV decoupling).
}
\end{figure}

In a pure EDMFT scheme, steps 5 (a, b, c) are skipped. Note that the form of the GW self-energy, $G W^c$, is chosen such that no double counting of the Hartree energy occurs. Figure~\ref{fig:Computational-scheme} gives an overview of the implementation of the GW+DMFT scheme.

\section{Results\label{results}}

In this section, we present numerical results for the $U$-$V$ Hubbard model using the different approximate formalisms discussed in the previous sections. We solve the impurity problems using the CTQMC-hyb method. Unless otherwise stated, the computations are performed at inverse temperature $\beta D=100$. The $k$-sums are discretized in the irreducible Brillouin zone on a $80\times80$ grid, while the imaginary time correlation functions are measured on a grid of $N=1000$ equally spaced points. Up to 40 EDMFT steps are required to reach convergence close to the Mott transition.

\subsection{Phase diagram\label{phase_diagram}}

\begin{figure}[h]
\begin{centering}
\includegraphics[scale=0.6]{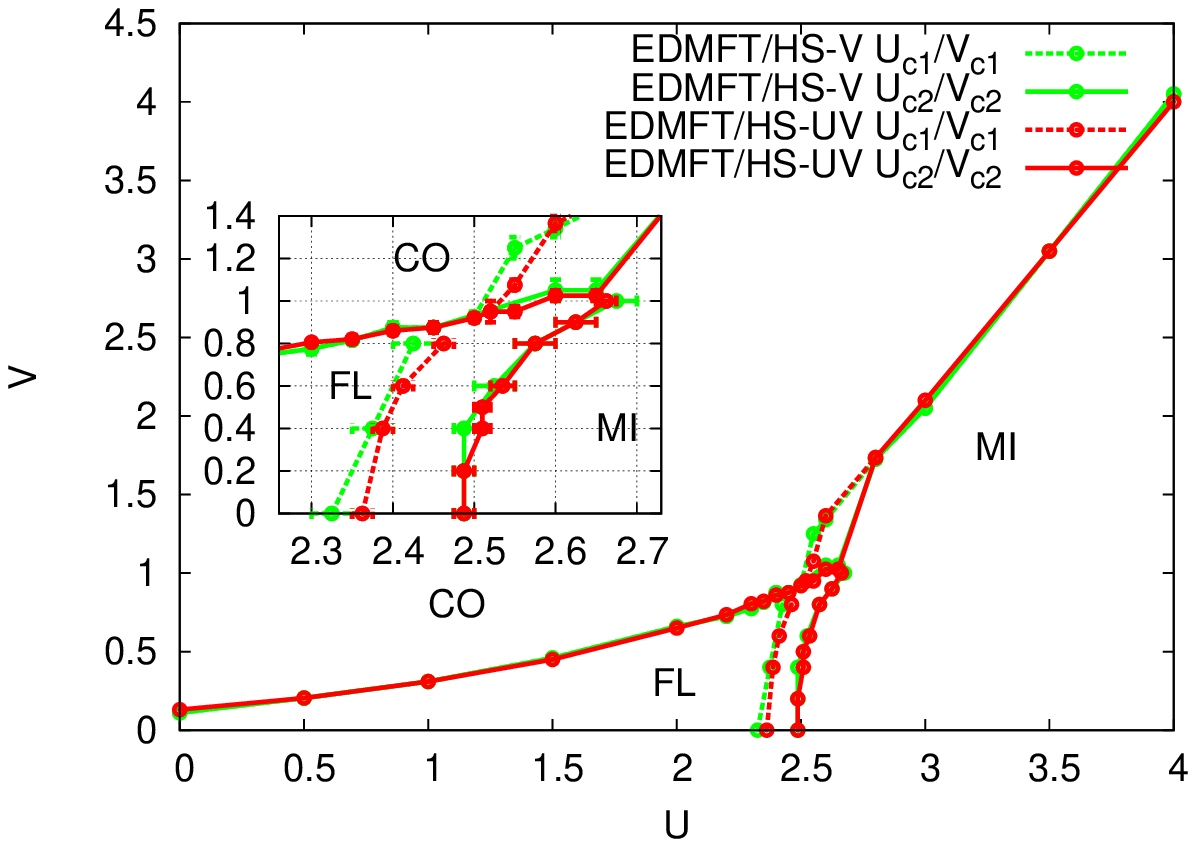}
\includegraphics[scale=0.6]{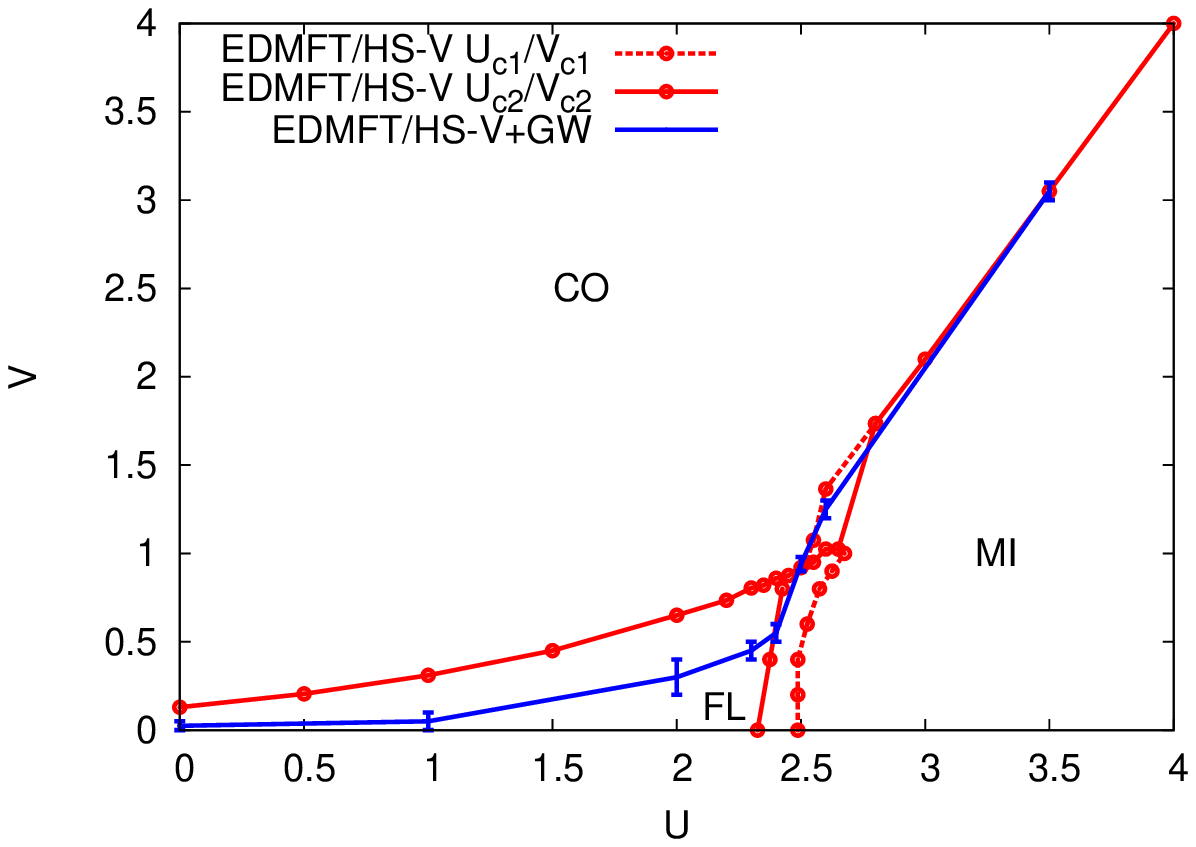}
\end{centering}
\caption{\label{fig:phase-diagram}
Phase diagram of the $U$-$V$ Hubbard model for HS-UV (red), and HS-V (green) decoupling. The top panel show the EDMFT result, 
the bottom panel the GW+EDMFT result. 
}
\end{figure}

Figure~\ref{fig:phase-diagram} shows the phase diagram in the space of the parameters 
$U$ and $V$ for the two decoupling schemes HS-UV and HS-V. The top panel shows the EDMFT result, the bottom panel
corresponds to GW+DMFT. 
There are three phases: (i) a Fermi liquid (FL) metal at small $U$ and small $V$, (ii) a charge-ordered (CO) insulator
at small $U$ and large $V$, and (iii) a Mott insulating (MI) phase at large $U$ and small $V$. 

The phase boundary to the charge-ordered phase has been located by approaching the phase transition from below $V_c$. The phase transition corresponds to a diverging charge susceptibility, namely to the formation of a pole in the Fourier transform $\chi(k,\omega)$ of $\chi_{ij}(t-t')\equiv\partial \langle n_i(t) \rangle / \partial h_j(t')$, where $h_j(t)$ is a probe field. Specifically, the charge-ordering transition will be signaled by a divergence at $Q=(\pi,\pi)$ and $\omega=0$ since the probe field for this phase is $h_i(t) = h e^{i Q R_i}$.  Using the action~(\ref{eq:U-V_action}), one can easily show that $\chi_{ij}(t-t')=\langle\bar{n}_i(t)\bar{n}_j(t')\rangle$. Recalling that $W = v - v\chi v$, we find the exact relation 
\begin{equation}
\chi(k,\omega) = -\frac {\Pi(k,\omega)}{1-v_k \Pi(k,\omega)}
\label{eq:criterion_susc}
\end{equation}
for the HS-UV scheme. Similarly, for the HS-V scheme, $\chi_V$ can be computed from $D=v^{nl}-v^{nl} \chi_V v^{nl}$ or 
$\chi_V=-\Pi_V/(1-v^{nl}\Pi_V)$. This shows that the transition also corresponds to the appearance of a pole in the fully-screened interaction $W(k,i\nu_n)$, and provides a rigorous definition of $V_c$ for HS-UV and HS-V, respectively: 
\begin{eqnarray}
1-(U-4V_c)\Pi\left[k=(\pi,\pi),\omega=0\right] & = & 0, \\
1+4 V_c \Pi_V\left[k=(\pi,\pi),\omega=0\right] & = & 0.
\label{eq:instability_criterion}
\end{eqnarray}
 
On the other hand, the phase boundary between the metal and the Mott insulator is signaled by a vanishing spectral weight at the Fermi level, which is related to the imaginary-time Green's function by $A(\omega =0)= \lim_{\beta\rightarrow\infty}\frac{\beta}{2} G(\frac{\beta}{2})$. 
The curvature of the FL-MI phase boundary shows that increasing the nearest-neighbor repulsion $V$ makes the system more metallic.

Within EDMFT, both decoupling schemes yield very similar phase diagrams. The boundary of the charge-ordered phase is characterized by two main regimes: for $U<U_c\approx 2.5$, $dV_c/dU\approx 1/4$, which is the prediction of mean-field studies. For $U>U_c$, $dV_c/dU\approx 2$. The transition between the two regimes is marked by a kink.  This kink also coincides with the point where the charge-ordered critical line meets the Mott critical line $U_c(V)$. The latter is only weakly dependent on $V$. In the temperature range $\beta D \in [25,100]$, the phase diagram does not depend much on temperature. 
The critical value $V_c(U)$ for $U<U_c$ is substantially larger than its naive mean-field estimate. We note, however, that EDMFT may overestimate the effect of the local interaction, leaving the true value of $V_c$ at an intermediate value.

The effect of the GW contribution to the phase diagram depends on the decoupling scheme. For HS-UV, GW does not have any influence of the phase boundaries, while in HS-V, GW substantially lowers the FL-CO phase boundary. 
This has the following origin: the HS-V scheme resums the diagrammatic series for $V$ and for $U$ separately (and treats $U$ only to second order), whereas the HS-UV scheme resums both terms simultaneously. HS-UV is thus better poised to capture the competition between the localizing term $U$ and the delocalizing term $V$. That GW in this scheme does not alter the phase boundaries should therefore come as no surprise: it merely shows that the local (EDMFT) physics alone fixes the critical value of the non-local interaction, and suggests that the HS-V decoupling underestimates $V_c$. For this reason, we will henceforth restrict most of our attention to the HS-UV scheme.

\begin{figure}[h]
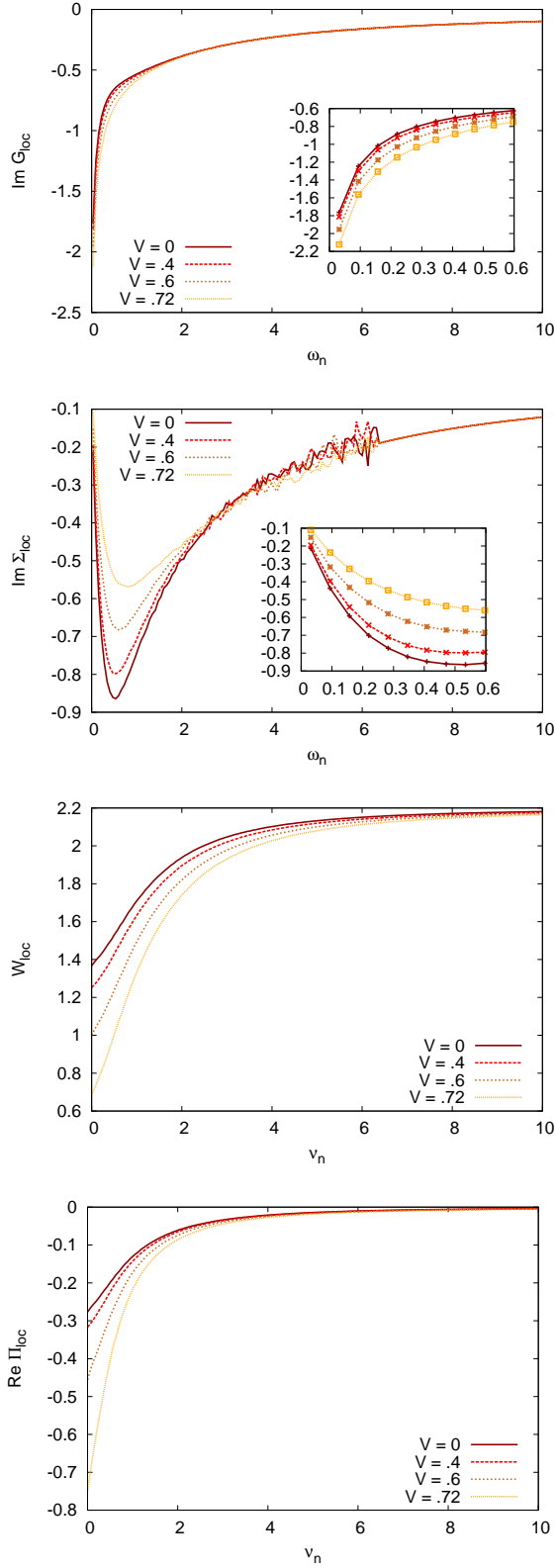

\begin{centering}
\includegraphics[scale=0.6]{{{G_loc_U_2.2}}}\\
\includegraphics[scale=0.6]{{{Sigma_imp_up_U_2.2}}}\\ 
\includegraphics[scale=0.6]{{{W_loc_U_2.2}}}\\
\includegraphics[scale=0.6]{{{P_imp_U_2.2}}}
\end{centering}
\caption{\label{fig:imaginary_time_observables}
Imaginary-frequency data for the EDMFT calculations at $U=2.2$ and indicated values of $V$ (HS-UV scheme). Panel (a):
$\mathrm{Im}G_\text{loc}(i\omega_{n})$. Panel (b): $\mathrm{Im}\Sigma_\text{loc}(i\omega_{n})$. 
Panel (c): $\mathrm{Re}W_\text{loc}(i\nu_{n})$. Panel (d): $\mathrm{Re}\Pi_\text{loc}(i\nu_{n})$. 
}

\end{figure}

Fig.~\ref{fig:imaginary_time_observables} plots the results for $\text{Im}G_\text{loc}(i\omega_n)$, $\text{Re} W_\text{loc}(i\omega_n)$,
$\text{Im}\Sigma_\text{loc}(i\omega_n)$ and $\text{Re}\Pi_\text{loc}(i\omega_n)$ corresponding to the EDMFT simulation at $U=2.2$      
and various values of $V$. As $V$ grows, $|\mathrm{Im}G(i\omega_{0})|$ increases and $|\Sigma_\text{loc}(i\omega_{0})|$
decreases, which indicates that the system becomes more metallic as a
result of screening by $V$. Indeed, the screening effect can be quantified by the static values of the fully screened interaction $W(0)\equiv W_\text{loc}(i\nu_{0})$ and of the partially screened interaction $U(0)\equiv \mathcal{U}(i\nu_{0})$, which are plotted in Fig.~\ref{fig:W0}. Nearest-neighbor repulsion $V$ induces a screening of the on-site Hubbard $U$, which becomes $U(0)$. 
When $V$ increases, $W(0)$ and $U(0)$ get smaller and smaller, resulting in a more metallic behavior. The transition
to the charge ordered insulator occurs around when the cost of doubloon formation vanishes, \emph{i.e.} when $W(0)=0$.

The last panel of Fig.~\ref{fig:imaginary_time_observables} shows the polarization $\Pi_\text{loc}(i\nu_n)$ (which is the local bosonic self-energy). $|\Pi_\text{loc}(i\nu_0)|$ gets larger as one approaches the phase boundary.

\clearpage

\subsection{Screening\label{screening}}

\subsubsection{Screened interaction and screening frequency}

\begin{figure}[h]
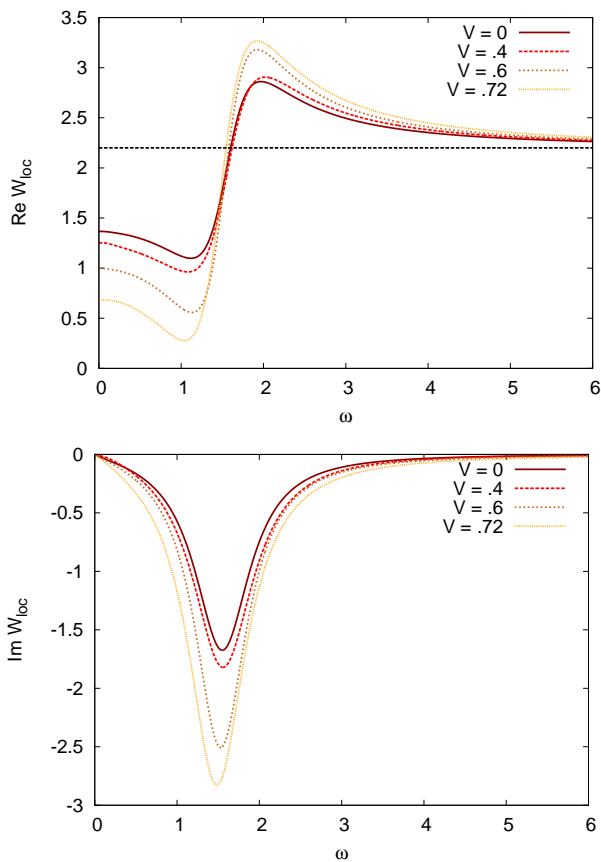

\begin{centering}
\includegraphics[scale=0.65]{{{W_loc_omega_U_2.2_Re}}}
\includegraphics[scale=0.65]{{{W_loc_omega_U_2.2_Im}}}
\end{centering}
\caption{\label{fig:W_omega}
Influence of $V$ on the effective local interaction for $U=2.2$ (HS-UV scheme).
\emph{Upper panel}: $\mathrm{Re}W_{\mathrm{loc}}(\omega)$.
\emph{Lower panel}: $\mathrm{Im}W_{\mathrm{loc}}(\omega)$. 
}
\end{figure}

The off-site interactions translate into an effective retarded interaction at the level of the impurity action, as made apparent in Eq.~(\ref{eq:Effective_action_1}). 
The frequency dependent local
interactions in the HS-UV formalism are now described by $\mathcal{U}(\omega)$ (or $U+\mathcal{D}(\omega)$ in the HS-V decoupling scheme). An important observable is $W_\text{loc}$, which is the fully-screened interaction, as opposed to $\mathcal{U}$, which excludes local screening effects. We have decided to focus more specifically on $W_\text{loc}$, which we have analytically continued to real frequencies using a Pad\'e scheme.\cite{Vidberg_1977} 
The shape of $W_\text{loc}(\omega)$ at $U=2.2$ and various $V$ is displayed
in Fig.~\ref{fig:W_omega}. $W_\text{loc}(\omega)$ has the typical shape of a screened interaction: the real part features two distinct energy scales, a
bare interaction $W_{\infty}=W_\text{loc}(\omega\rightarrow\infty)=U$
at high energies and a screened interaction $W(0)=W_\text{loc}(\omega=0)<U$ at low energies, separated by a screening frequency $\omega_0$. Its Kramers-Kronig-conjugated imaginary part has most of its spectral weight concentrated around $\omega_0$.

In the following, we will mainly focus on three parameters to describe screening effects: (i) the value of the static screened interaction $W(0)$, (ii) the strength $\lambda$ of this screening, which we will define later, and (iii) the screening frequency $\omega_0$.

\begin{figure}[h]
\begin{centering}
\includegraphics[scale=0.65]{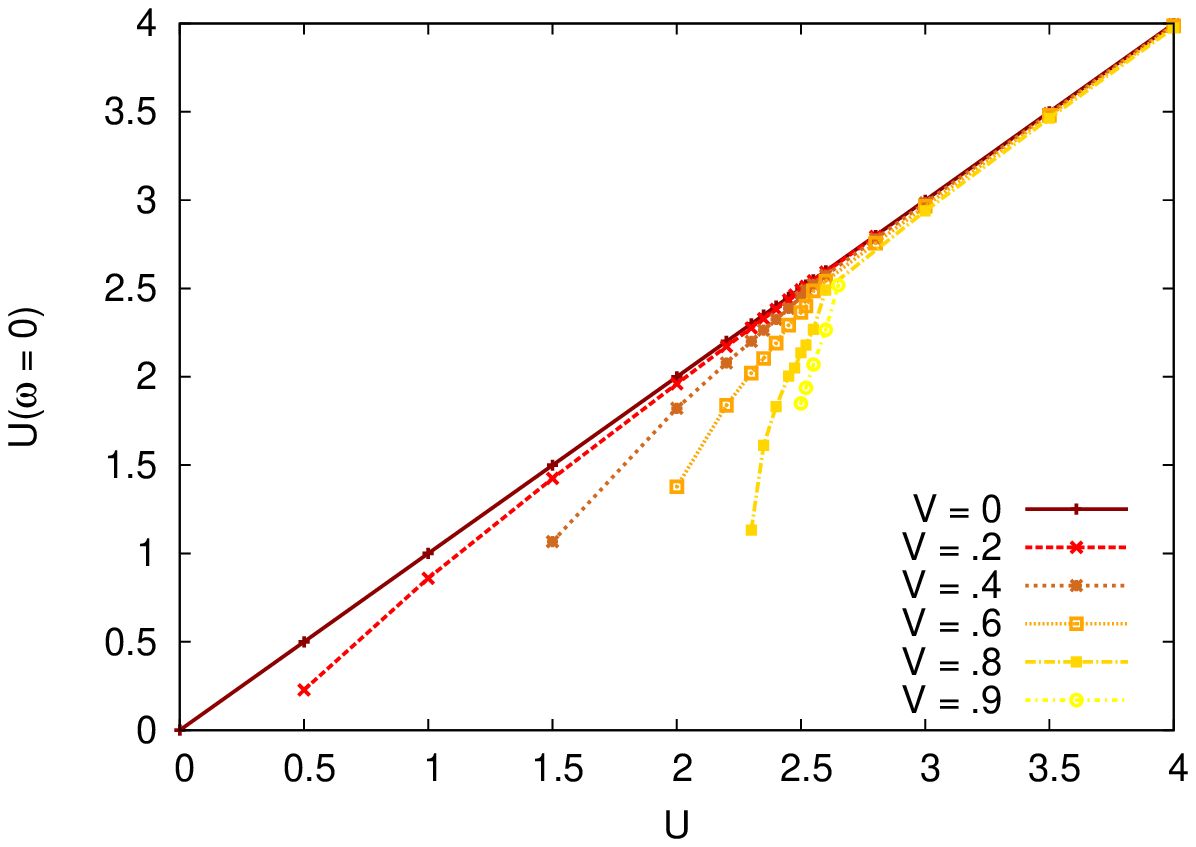}
\includegraphics[scale=0.65]{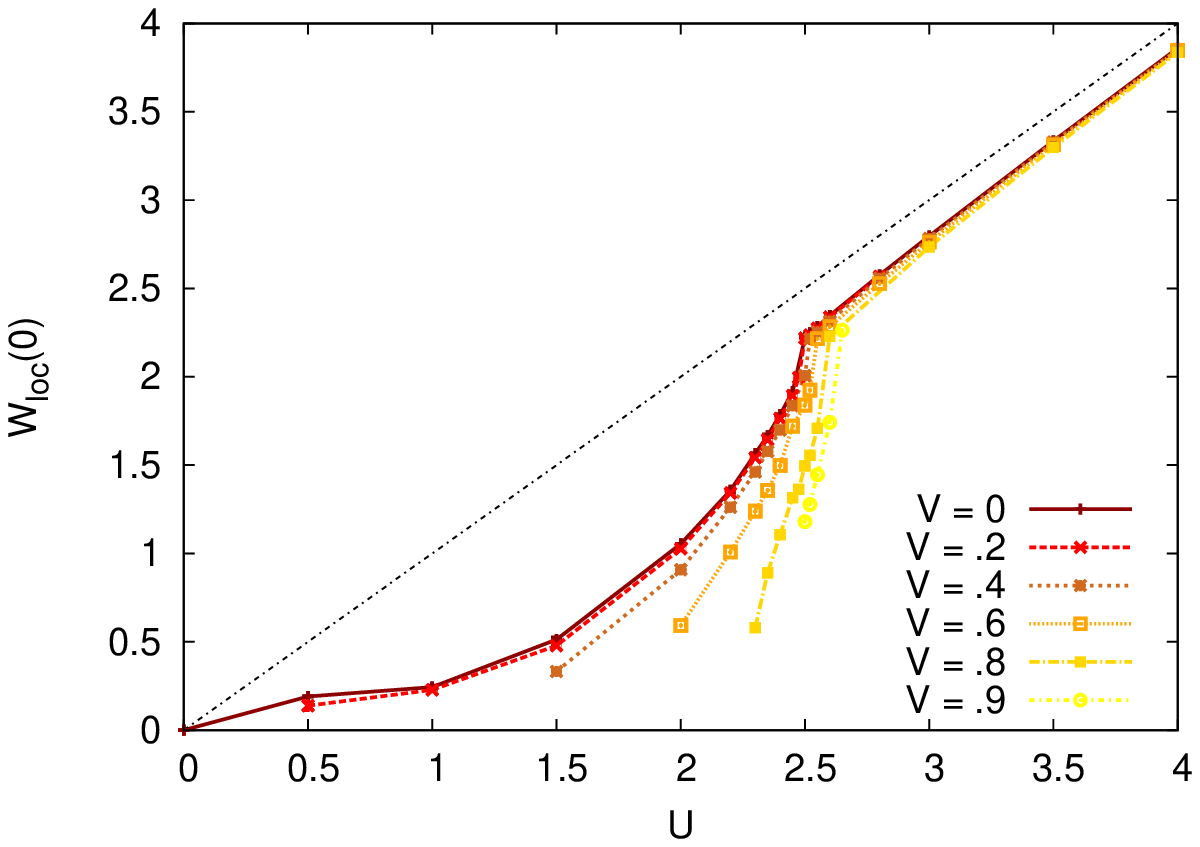}
\includegraphics[scale=0.65]{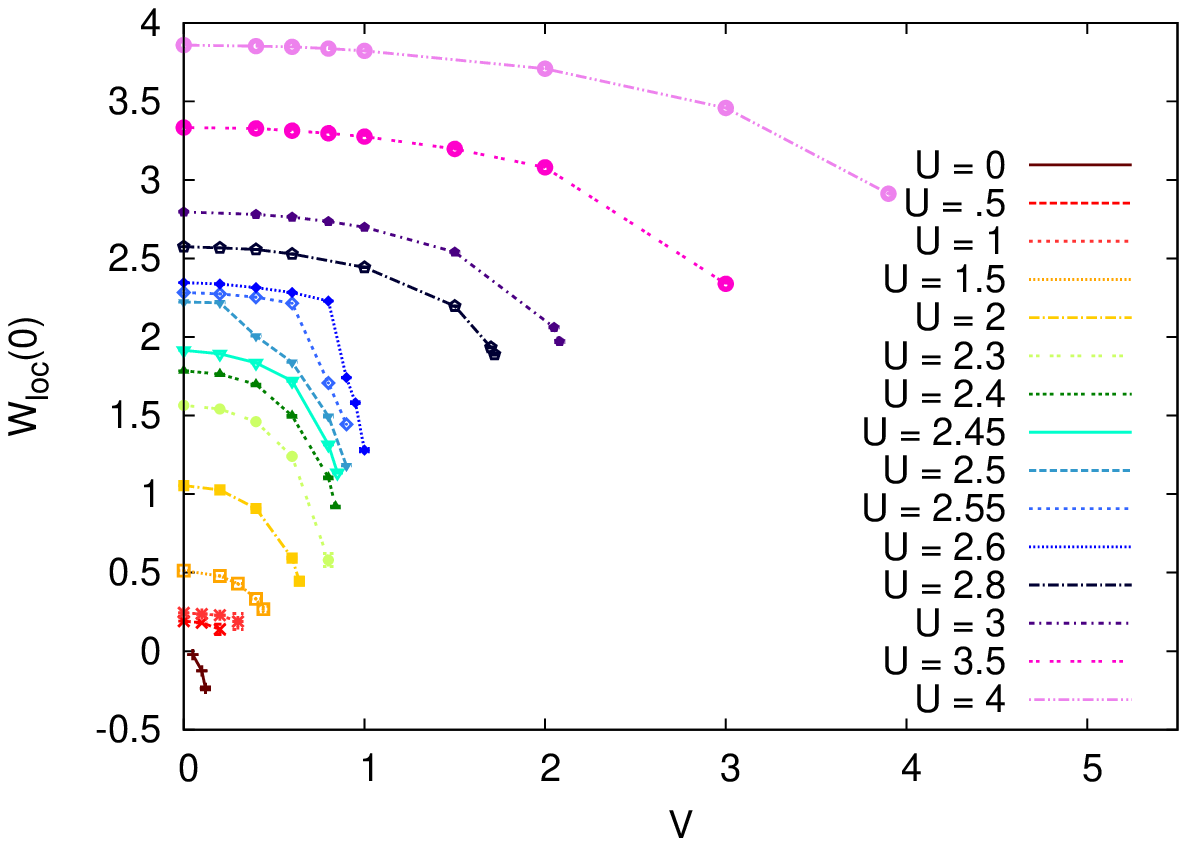}
\end{centering}
\caption{\label{fig:W0}
Static screened interactions in the HS-UV scheme. \emph{Upper panel}: partially screened $\mathcal{U}(\omega=0)$ as a function of $U$. \emph{Middle panel}: fully screened $W_\text{loc}(\omega=0)$ as a function of $U$. \emph{Lower panel}: $W_\text{loc}(\omega=0)$ as a function of $V$.
}
\end{figure}

$W(0)$ is the effective fully-screened interaction between two electrons on the same lattice site. Its evolution across the $U$-$V$ plane for the HS-UV scheme is illustrated in the middle and lower panels of Fig.~\ref{fig:W0}. $W(0)$ decreases for increasing $V$, and drops to zero as $V$ approaches $V_c$. This is intuitively easy to understand: the critical line corresponds to the locus where the cost $W_\text{loc}(\omega=0)$ for the formation of doublons vanishes. The lower panel of Fig.~\ref{fig:W0}) shows that the screening of the local interaction is much more efficient in the metallic phase than in the Mott insulator. While the screening is especially large in the metallic phase, it is weak and weakly $V$-dependent in the Mott insulating phase (at least for low to intermediate values of $V$). 

Let us emphasize that there are screening effects even when $V = 0$ in the metallic phase (middle panel, red curve with crosses). This shows that in a EDMFT description of the simple Hubbard model ($V=0$), there is a screening of the static $U$ by the \emph{local} polarization caused by $U$ itself, provided one uses the HS-UV decoupling scheme. In the HS-V method, the screening comes only from $V$, as $D$ originates from the HS decoupling of the nearest-neighbor interaction only. The upper panel of Fig.~\ref{fig:W0} shows $\mathcal{U}(\omega=0)$ for comparison. This corresponds to the local static interactions without polarization effects. As expected, $\mathcal{U}(\omega=0)>W(0)$ since the local polarization further screens the local interaction.

Motivated by the Hamiltonian representation of the impurity model with dynamically screened interaction (see Appendix \ref{hamiltonian_formulation}) we define the strength of the screening by the parameter
 $\lambda\equiv\sqrt{|\int_{0}^{\infty}d\omega\mathrm{Im}\mathcal{U}(\omega)|}$. It follows from Eq.~(\ref{eq:Im_W_imp}) that $\lambda\propto\sqrt{\sum_{p}\lambda_{p}^{2}}$, where $\lambda_{p}$ is the coupling of the harmonic oscillator with frequency $\omega_{p}$ to the charge on the impurity, $n_{0}$. Therefore, $\lambda$ can be interpreted as the strength of the coupling to the charge fluctuations. Its dependence on $U$ and $V$ is presented in the upper panel of Fig.~\ref{fig:Omega_0}. Except in the vicinity of the phase transition to the charge ordered phase, $\lambda$ is proportional
to $V$. The square of the proportionality constant (see inset) decreases with increasing $U$,
and exhibits two regimes separated by a kink at $U=U_c$: for $U\leq U_{c}$, $\frac{d}{dU}\left[\left(\frac{d\lambda}{dV}\right)^2\right]\approx-2.2$, otherwise
$\frac{d}{dU}\left[\left(\frac{d\lambda}{dV}\right)^2\right]\approx-0.34$. This again indicates that screening in the Mott insulator is radically different from screening in the Fermi-liquid metal.

A relevant question is in which parameter regime a model with a static screened interaction provides a reasonable approximation. 
A useful quantity to investigate in this context is the screening frequency $\omega_0$, whose precise determination is a somewhat tricky task owing to the Pad\'e procedure's inaccuracy. Instead of measuring $\omega_0$ as the minimum of $\mathrm{Im}W_\text{loc}(\omega)$, we have found it more meaningful to determine it from the first moment of $\mathrm{Im}W_\text{loc}(\omega)$: $\omega_0 \approx \langle \omega \rangle \equiv \int_0^{\infty} d\omega \omega \mathrm{Im}W_\text{loc}(\omega)/\int_0^{\infty} d\omega \mathrm{Im}W_\text{loc}(\omega)$, whose dependence on $U$ and
$V$ in the HS-UV scheme is presented in Fig.~\ref{fig:Omega_0} (middle panel).
This figure shows that the screening frequency is only weakly dependent on the nearest-neighbor interaction $V$. On the other hand, the larger the bare interaction $U=U_{\infty}$, the larger the screening frequency. 
The lower panel shows that the screening frequency (for $V=0$) has two distinct regimes separated by a kink at $U=U_c$.

\subsubsection{Screening frequency from linearized DMFT}

The $U$-dependence of the screening frequency may be traced back to the  form of the one-particle local spectrum. As can be seen from Eq.~(\ref{eq:W_Chi_relation}), for $V = 0$,\footnote{In this case, one can check that within EDMFT, $\mathcal{U}=U$.} the frequency-dependence of $W_{\text{loc}}$ (and thus the value of $\omega_0$) is inherited from the charge-charge correlation function $\chi_{\text{loc}}$. An analytical estimate for the poles of this function can be calculated by means of a simple approximation named linearized DMFT.\cite{Potthoff_2001}
In this method, the impurity problem is approximated by the coupling of the correlated impurity to a \emph{single} uncorrelated bath level describing the hybridization of the impurity to the lattice degrees of freedom. This simplified version of the impurity problem allows for the explicit calculation of the local Green's function:\cite{Lange_1998}

\begin{equation}
G_{\text{loc}}^{l-\mathrm{DMFT}}(\omega)=\sum_{i=1}^{2}a_{i}\left\{ \frac{1}{\omega-\bar{\epsilon}_{i}}+\frac{1}{\omega+\bar{\epsilon}_{i}}\right\} 
\end{equation}
with 
\begin{eqnarray}
\bar{\epsilon}_{\substack{1\\
2
}
} & = & \frac{1}{4}\left(\sqrt{U^{2}+64V_\text{hyb}^{2}}\mp\sqrt{U^{2}+16V_\text{hyb}^{2}}\right),\\
a_{1} & = & \frac{1}{4}\left(1-\frac{U^{2}-32V_\text{hyb}^{2}}{\sqrt{U^{2}+64V_\text{hyb}^{2}}\sqrt{U^{2}+16V_\text{hyb}^{2}}}\right),
\end{eqnarray}
as well as $a_{2}=\frac{1}{2}-a_{1}$. The hybridization strength's
dependence on $U$ is given by $V_\text{hyb}=t\sqrt{z}\sqrt{1-U^{2}/U_{c}^{2}}$
(see Ref. \onlinecite{Potthoff_2001} for details). $U_{c}$ denotes the
critical $U$ for the Mott transition. In the estimate below, we will
use the value computed within EDMFT, $U_{c}=2.5$.

In the absence of vertex corrections, the corresponding charge-charge correlations function can be computed as $\chi_{\text{loc}} = - 2 G_{\text{loc}} G_{\text{loc}}$, leading to the expression
\begin{equation}
\chi_{\text{loc}}(\omega)=-2\bigg\{\frac{2a_{1}^{2}\epsilon_{1}}{\omega^{2}-\epsilon_{1}^{2}}+\frac{2a_{2}^{2}\epsilon_{2}}{\omega^{2}-\epsilon_{2}^{2}}+\frac{4a_{1}a_{2}\epsilon_{3}}{\omega^{2}-\epsilon_{3}^{2}}\bigg\}\label{eq:chi_imp_lin_DMFT}
\end{equation}
The six ($3\times2$) poles are defined as $\epsilon_{1}=2\bar{\epsilon}_{1}$,
$\epsilon_{2}=2\bar{\epsilon}_{2}$ and $\epsilon_{3}=\bar{\epsilon}_{1}+\bar{\epsilon}_{2}$.

\begin{figure}[h!]
\begin{centering}
\includegraphics[scale=0.64]{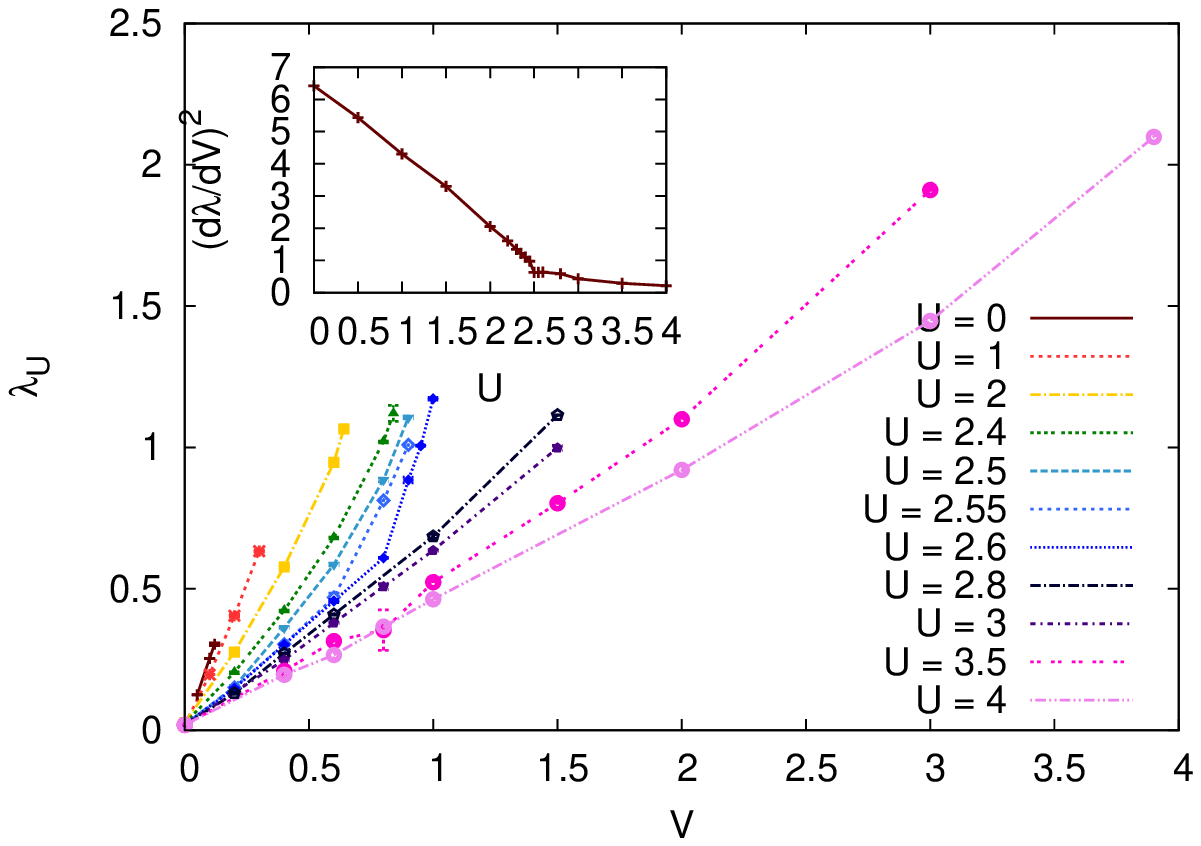}
\includegraphics[scale=0.64]{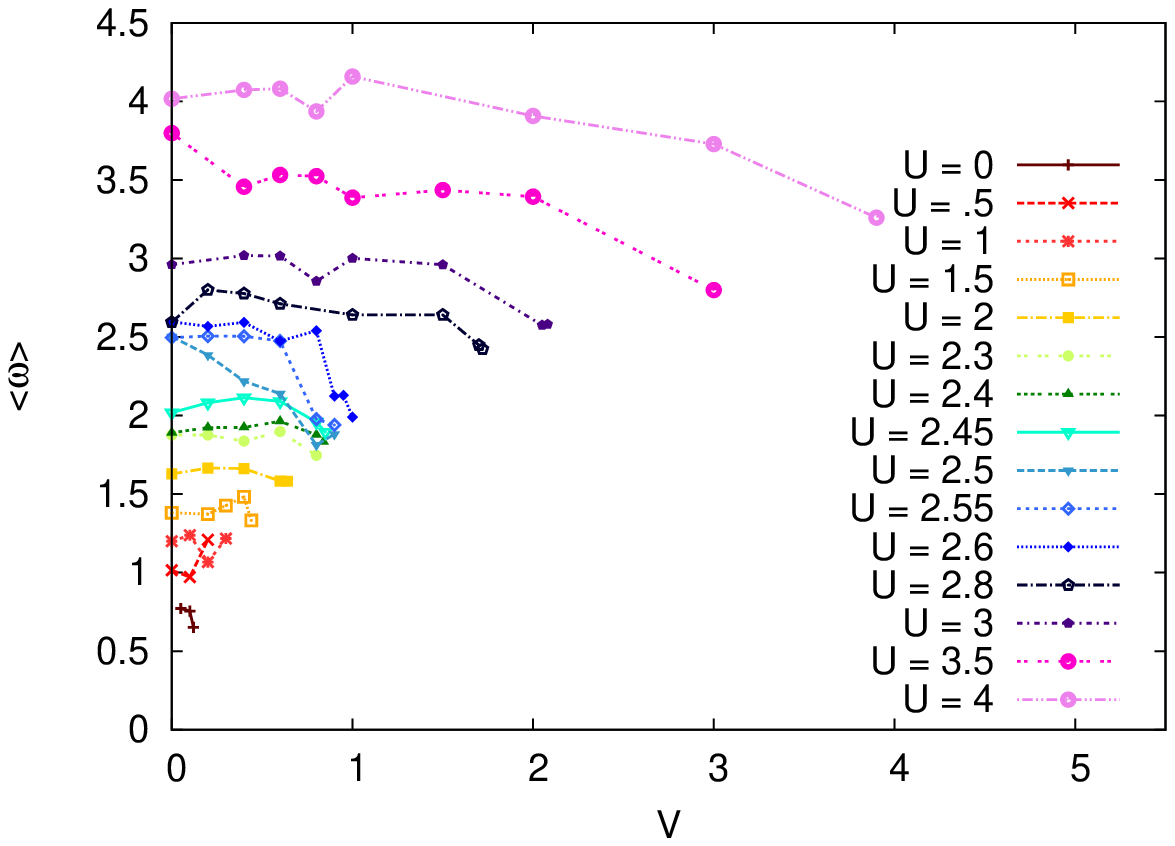}
\includegraphics[scale=0.44]{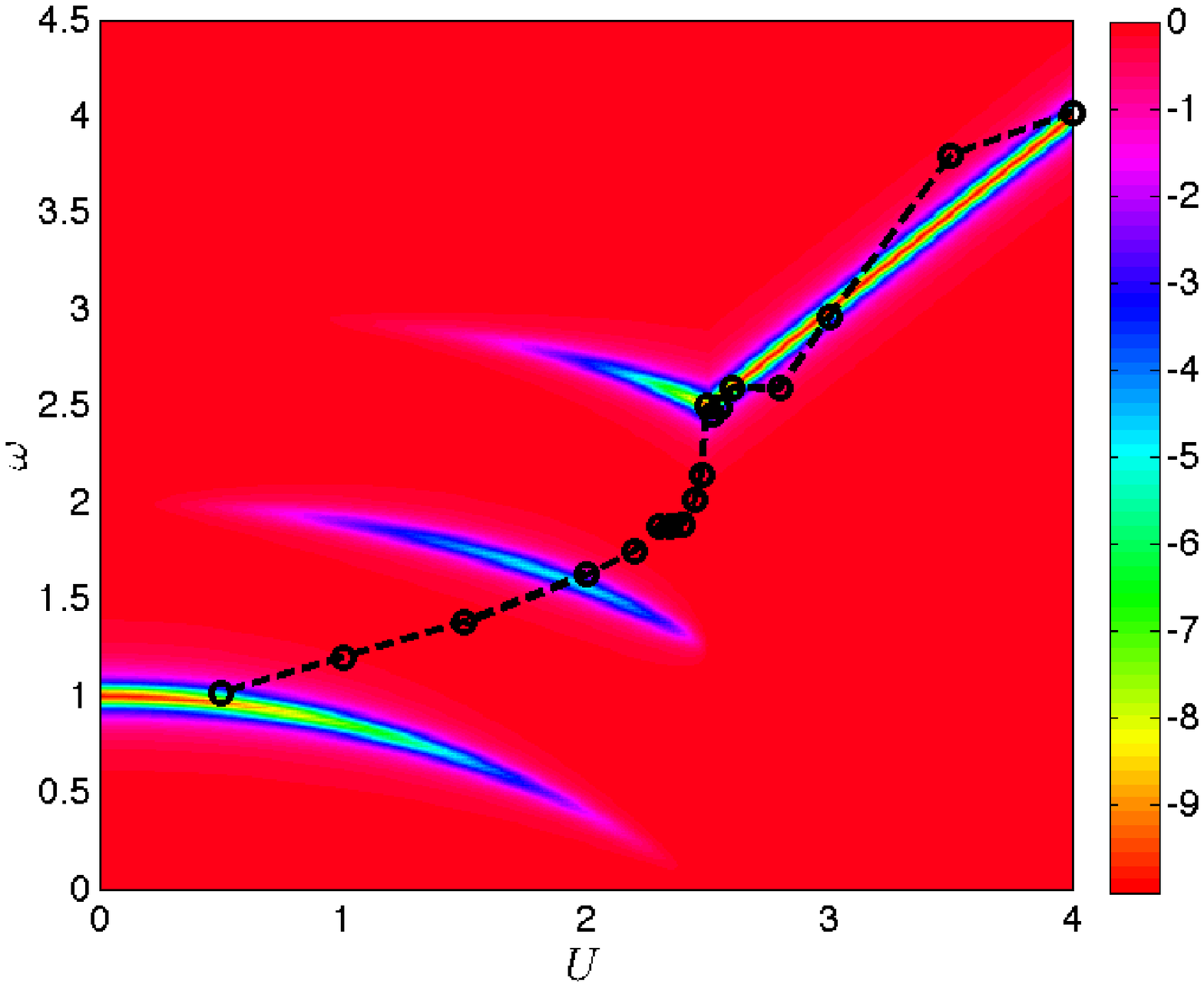}
\end{centering}
\caption{\label{fig:Omega_0}
Screening parameters.
\emph{Upper panel}: $\lambda$ as a function of $V$ (HS-UV scheme). \emph{Inset}: $(\frac{d\lambda}{dV})^2$ as a function of $U$.
\emph{Middle panel}: $\langle\omega\rangle$ as a function of $V$.
\emph{Lower panel}: $\text{Im}\chi_{\text{loc}}^{l-DMFT}$ in the ($\omega$,$U$) plane for $V=0$. The poles have been artificially broadened by an imaginary factor $\eta=0.01$. Black line: EDMFT result for $\langle\omega\rangle$ as a function of $U$ (for $V=0$). 
}
\end{figure}

These poles, displayed in the lower panel of Fig.~\ref{fig:Omega_0}, correspond to the transitions allowed in the various correlation regimes, namely: in the low-correlation limit, only transitions within the quasi-particle peak are possible. As correlations increase, the appearance of Hubbard bands enable additional transitions from the lower Hubbard band to the unoccupied states of the quasi-particle peak, and from the occupied states of the quasi-particle peak to the upper Hubbard band. In the strong correlation regime, finally, the only possible transitions are those between the lower and the upper Hubbard band.

\subsection{Spectral Properties}

\begin{figure}[t!]
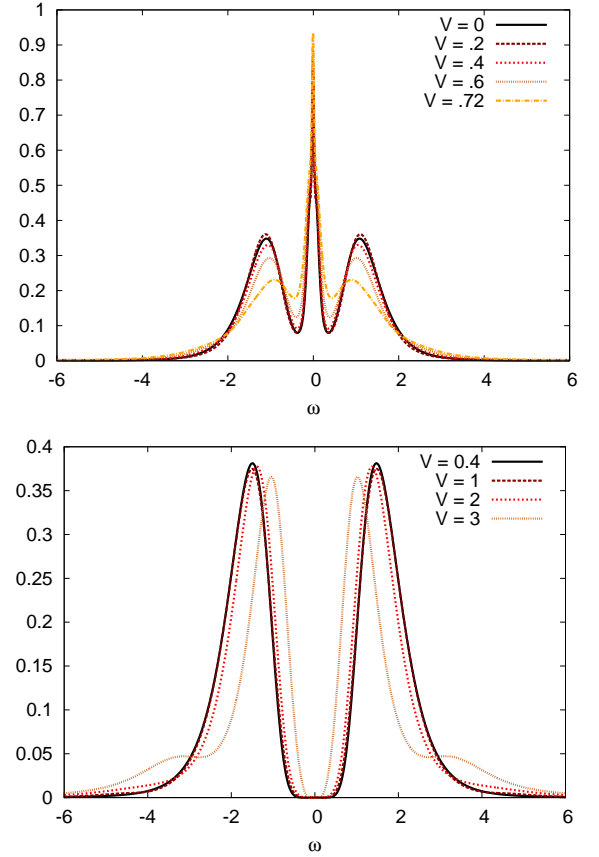

\begin{centering}
\includegraphics[scale=0.65]{{{spectra_DALA_U_2.2}}}
\includegraphics[scale=0.65]{{{spectra_DALA_U_3.5}}}
\end{centering}
\caption{\label{fig:Spectral_function}Local spectral function (HS-UV scheme). \emph{Upper panel}: $U=2.2$. \emph{Lower panel}: $U=3.5$. 
}

\end{figure}

The screening effects coming from $U$ and $V$ have some impact on the local spectral function. In the weakly-correlated regime ($U<U_c$), the non-local interactions $V$ tend to smooth out the incoherent Hubbard bands and transfer some spectral weight to the quasi-particle peak and into the gap region between the quasi-particle peak and the Hubbard bands (Fig.~\ref{fig:Spectral_function}, upper panel). This behavior is consistent with the imaginary-frequency data showing a more metallic behavior as $V$ increases. The effects are more dramatic in the strongly correlated regime ($U>U_c$), where one observes new features in the local spectral function, as shown in the lower panel of Fig.~\ref{fig:Spectral_function}. In addition to the two Hubbard bands located at $\omega=\pm U/2$,
the spectral function has two symmetric satellites at $\omega=\pm(U/2+\omega_{0})$,
whose spectral weight grows with $V$. The position of the peaks comes
from the convolution in Eq.~(\ref{eq:Convolution}), since $A_\text{aux}$
contains spectral weight at $\pm U/2$ and $B$ contains weight at
$\pm\omega_{0}$.

\subsection{Momentum dependence}

We now turn to the analysis of the GW+EDMFT results. 

\subsubsection{Non-local self-energies}

\begin{figure}[ht]
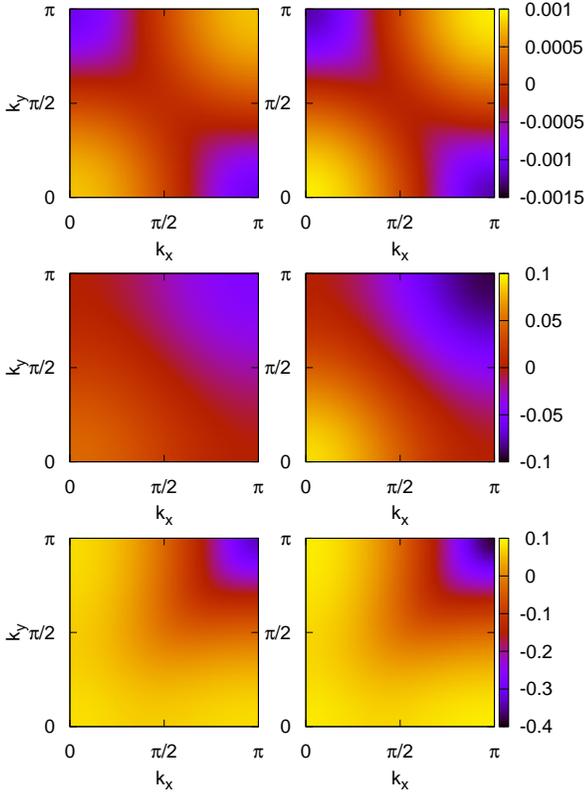
 

\includegraphics[scale=0.65,trim=0 100 0 50]{{{Sigma_GW_nonlocal_up_U_2_V_.4_V_.64_Im}}}
\includegraphics[scale=0.65,trim=0 100 0 0]{{{Sigma_GW_nonlocal_up_U_2_V_.4_V_.64_Re}}}
\includegraphics[scale=0.65,trim=0 50 0 0]{{{P_GW_nonlocal_U_2_V_.4_V_.64}}}

\caption{\label{fig:GW_nonlocal}
GW Non-local contribution in the HS-UV scheme at $U = 2$ and $V = .4$ (left) or $V = .64$ (right). \emph{Upper panel}: $\mathrm{Im}\Sigma_{nonloc}^{GW}(k,i\omega_0)$. \emph{Middle panel}: $\mathrm{Re}\Sigma_{nonloc}^{GW}(k,i\omega_0)$. \emph{Lower panel}: $\mathrm{Re}\Pi_{nonloc}^{GW}(k,i\nu_0)$.
}

\end{figure}

Figure~\ref{fig:GW_nonlocal} displays $\Sigma_\text{nonlocal}^{GW}(k,i\omega_0)$ and $\Pi_\text{nonlocal}^{GW}(k,i\nu_0)$ in the metallic phase near the charge ordering transition. These quantities vanish in the limit of large dimensions and are thus neglected in the EDMFT treatment.  The GW contribution to the imaginary part of the electronic self-energy $\Sigma$ is negligible with respect to the local self-energy (for instance, at $U=2$ and $V=0.6$, $\mathrm{Im}\Sigma_\text{loc}(i\omega_0)=-0.18$, compared to a non-local GW self-energy $<0.001$). This holds across the Fermi-liquid phase and the Mott insulating phase. The real part of $\Sigma^{GW}_\text{nonlocal}$ is relatively large away from the EDMFT Fermi surface, but does not alter this Fermi surface.  On the other hand, the non-local polarization is comparable to its local counterpart ($\Pi_\text{loc}(i\nu_0)=-0.39$ for $U=2$, $V=0.6$).

This should have an impact on the phase diagram if one recalls the criterion of Eq.~(\ref{eq:instability_criterion}). However, the \emph{local} observables are also modified in the self-consistent calculation, as will be described in the next subsection, which prevents a direct prediction of the effect of the non-local terms. As shown in Fig.~\ref{fig:phase-diagram}, the effect of GW depends on the decoupling scheme. For the HS-UV scheme, the GW contribution has a negligible influence on the phase diagram.  For the HS-V scheme, the GW contribution has a large effect on the phase boundary between the metallic phase and the charge-ordered phase. The non-local polarization, peaked at $k=(\pi,\pi)$, enhances nesting effects and leads to a substantially lower $V_c$ compared to EDMFT.

\subsubsection{Influence of the self-consistency on local observables}

\begin{figure}[ht] 
\centering{}\includegraphics[scale=0.65]{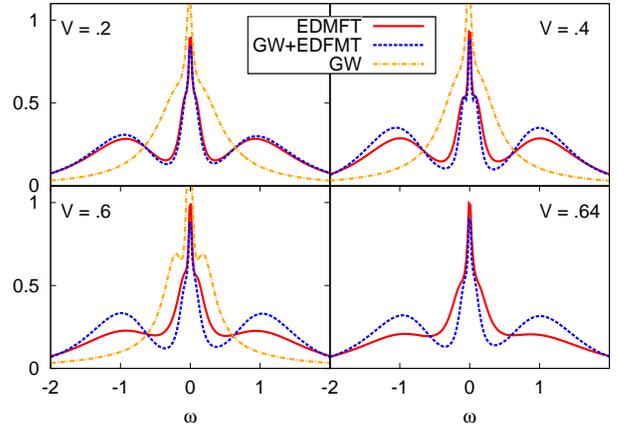}

\caption{\label{fig:GW_local}Effects of GW on the local spectral function $A_\text{loc}(\omega)$  (HS-UV scheme, $U=2$).}
\end{figure}

\begin{figure}[ht]
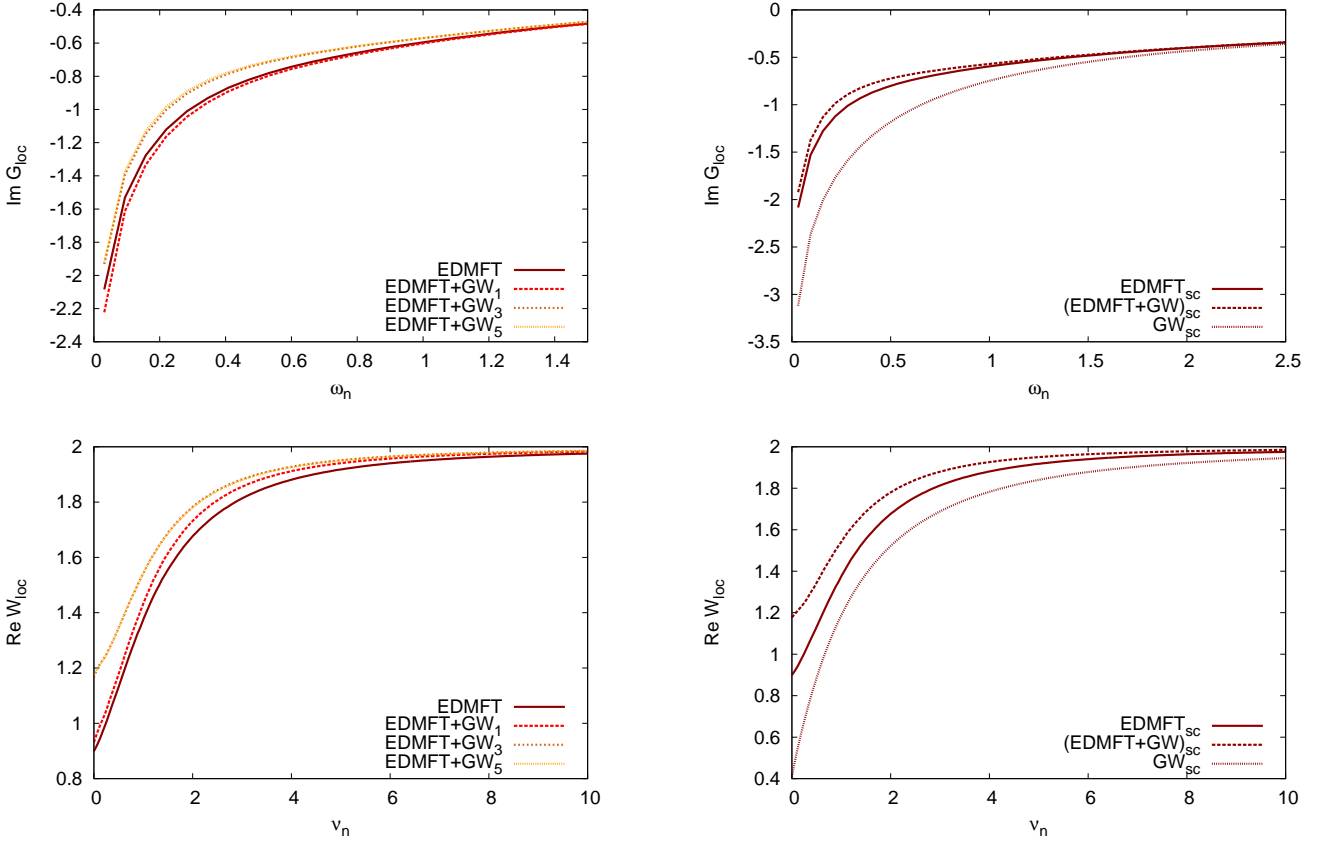
 

\centering{}\includegraphics[scale=0.65]{{{G_loc_U_2_V_.4_HSUV_EDMFT_EDMFTGW_convergence}}}
\centering{}\includegraphics[scale=0.65]{{{W_loc_U_2_V_.4_HSUV_EDMFT_EDMFTGW_convergence}}}

\caption{\label{fig:convergence_to_GW}Convergence properties for $U=2$, $V=0.4$ (HS-UV scheme). EDMFT corresponds to the converged EDMFT result, EDMFT+GW$_{i}$ denotes the observables corresponding the $i$th iteration with non-local self-energies, with EDMFT as starting input. \emph{Upper panel}: $\mathrm{Im}G_\text{loc}(i\omega_n)$.  \emph{Lower panel}: $\mathrm{Re}W_\text{loc}(i\nu_n)$. 
}
\end{figure}

As already hinted at, the self-consistency condition leads to an adjustment of the local quantities to the non-local self-energies. This is illustrated in Fig.~\ref{fig:GW_local}, which displays the local spectral function after EDMFT convergence and after GW+EDMFT convergence, and in Fig.~\ref{fig:convergence_to_GW}, which illustrates the convergence from EDMFT to EDMFT+GW.

The EDMFT+GW local spectral function features more pronounced Hubbard bands than the EDMFT spectrum. This is a sign of increased local correlations. Inspection of the imaginary frequency data corroborates this observation: $W_\text{loc}(\omega=0)$ is enhanced with respect to the EDMFT result, indicating that the local interactions are stronger. Likewise, $|\text{Im} G_\text{loc}(i\omega_0)|$ is reduced by GW.

These observations mean that the local quantities have become more ``insulating" in character as a result of the addition of the non-local GW self-energy. This can be interpreted in the following way: contrary to the EDMFT case, where all the screening and correlation effects are absorbed into the local self-energy, in EDMFT+GW some of these effects are now carried by the non-local components. Specifically, GW non-local self-energies carry important screening effects owing to the very nature of the GW approximation. This leads to a redistribution of the screening between local and non-local observables: local observables become less screened, and thus more correlated.

\subsubsection{Convergence properties and self-consistency}

\begin{figure}[ht]
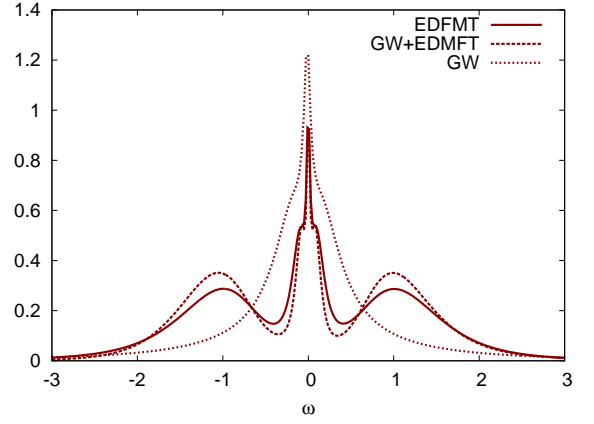


\centering{}\includegraphics[scale=0.65]{{{Im_G_loc_U_2_V_.4}}}
\centering{}\includegraphics[scale=0.65]{{{Re_W_loc_U_2_V_.4}}}
\centering{}\includegraphics[scale=0.65]{{{A_loc_U_2_V_.4_Maxent}}}

\caption{\label{fig:influence_self_consistency}Influence of the self-consistency for $U=2$, $V=0.4$ (HS-UV scheme). \emph{Upper panel}: $\mathrm{Im}G_\text{loc}(i\omega_n)$.  \emph{Middle panel}: $\mathrm{Re}W_\text{loc}(i\nu_n)$. \emph{Lower panel}: $A_\text{loc}(\omega)$ (obtained from MaxEnt continuation).}
\end{figure}

Figure~\ref{fig:influence_self_consistency} shows the converged $\text{Im}G_\text{loc}$ and $\text{Re} W_\text{loc}$ for the three self-consistent schemes: EDMFT alone, EDMFT+GW and GW alone (with the EDMFT+GW result as a starting point). As expected, GW is the most metallic in character and the corresponding spectrum does not have Hubbard bands.  Interestingly, GW+EDMFT is not a kind of ``average" between GW and EDMFT. It exhibits stronger correlation effects than both GW and EDMFT.

\section{Conclusions}

We have presented an application of 
the EDMFT and GW+DMFT methods to the single-band extended
Hubbard model.
In a first step, the two formalisms have been reviewed in detail:
We have 
discussed the construction of the free energy functional,
and compared two different flavors of such functionals that 
have been proposed in the literature,
corresponding to two distinct decouplings of the interaction term,
and leading respectively to the GW+DMFT and GD+SOPT+DMFT approaches.
We have presented the details of our implementation of a fully self-consistent
GW+DMFT scheme based on a numerically exact continuous-time quantum 
Monte-Carlo solver adapted for frequency-dependent local interactions. 
The investigation of the frequency-dependence of these interactions for 
parameters ranging from weak to strong coupling shows 
that the $U$-dependence of the local screening frequency reflects the 
form of the local one-particle spectrum.
We have investigated the spectral properties of the extended
Hubbard model within three self-consistent schemes, namely 
EDMFT, GW, and GW+DMFT.
The nearest-neighbor repulsion $V$ leads, in the Mott insulator, 
to high-energy satellites in the local spectra.

The GW+DMFT calculations demonstrate that the non-local contributions 
to the self-energy coming from the GW diagrams are quite small in the case of the extended Hubbard model. In 
view of the strong momentum dependence observed in self-energies obtained 
from cluster DMFT calculations for the two-dimensional Hubbard model
as one approaches the Mott transition,\cite{Werner_2009dca, Gull_2009}  our results
confirm the importance of spin fluctuations,
suggesting that further non-local 
diagrams have to be considered in order to capture the dominant fluctuations in the extended Hubbard model.

The model calculations presented in this paper can be 
straightforwardly extended to 
the multi-orbital case
and to additional, longer-range matrix elements of the screened 
interaction,
paving the way for realistic first-principles material calculations. 
It is worthwhile to note that the GW+DMFT method and its variations
are a computationally cheap way of incorporating the leading vertex 
contribution -- in the form of the EDMFT self-energy -- into the 
description of a solid, and to introduce some spatial fluctuations 
through a perturbative scheme. This contrasts with methods involving 
an explicit computation of the vertex functions 
\cite{Toschi_2007, Rubtsov_2008} whose implementation for simple 
model systems is already a formidable challenge.

In real materials, further degrees of freedom, 
stemming for example from the multi-band nature and ligand
orbitals, lead to a renormalisation
and/or frequency-dependence of the parameters in the low-energy description.
Relatively weak, but non-local correlation
effects are expected to play a dominant role in the case of extended ligand or
higher-lying empty states, thus providing an
additional motivation for a combination of GW and DMFT.
Indeed, GW provides an accurate and comparatively 
inexpensive description of the screening from ``uncorrelated'' bands, 
making the application of the GW+DMFT method to electronic 
structure calculations for realistic solids highly promising.

\bigskip{}

\acknowledgments{
We acknowledge useful discussions with 
F. Aryasetiawan, M. Casula,
A. Georges, M. Imada, A. Millis, and T. Miyake,
as well as computing time on the Brutus cluster at ETH Zurich.
This work was supported by the Swiss National Science Foundation (grant PP0022-118866), DFG FOR 1346 and by the French ANR under projects CORRELMAT and SURMOTT,
GENCI/IDRIS Orsay under project 1393.
The Monte-Carlo scheduler uses the ALPS library.\cite{ALPS}
}

\bigskip{}

\appendix

\section{Extended Hubbard models from the First Principles \label{remarks_extended_hubbard_models}}

The Hamiltonian $H$ of the extended Hubbard model is supposed
to provide the low-energy physics of correlated materials.
$H$ can be regarded as an effective Hamiltonian resulting from
a ``downfolding procedure'', which is expressed in some localized Wannier
basis. The downfolding procedure is akin to a renormalization group
transformation which, starting from all the bands resulting from
a LDA calculation, produces an effective model for the bands in a low-energy window 
by integrating out the remaining bands. In this process, the bare
Coulomb interaction $v(r,r')=4\pi e^{2}/|r-r'|$ is transformed into to a frequency-dependent
partially screened interaction $W_r(r,r',\omega)$,\cite{Aryasetiawan_2004,Miyake_2008}
which acts in the low-energy subspace. In principle, $W_r$ is computed as $W_r = v (1-v P_r)^{-1}$, where $P_r$ is the polarization computed when transitions within the effective model have been removed. The matrix elements of this interaction in the Wannier basis are
\begin{equation}
V_{ijkl}(\omega)=\int d^{3}rd^{3}r'\phi_{i}^{*}(r)\phi_{j}^{*}(r')W_r(r,r',\omega)\phi_{k}(r)\phi_{l}(r'),
\label{eq:Matrix_elements}
\end{equation}
where $\phi_{i}$ denotes a Wannier orbital centered at site $i$.

Model (\ref{eq:U-V_model}) involves three approximations on the above matrix
elements: (i) the frequency-dependence of $V_{ijkl}(\omega)$ is neglected:
$V_{ijkl}\equiv V_{ijkl}(\omega=0)$, 
(ii) the interaction is restricted to density-density terms 
$V_{ijkl}=V_{ijij}\delta_{ik}\delta_{jl}\equiv V_{ij}$, and 
(iii) only the on-site matrix element $U\equiv V_{ii}$ and
the nearest-neighbor matrix element $V\equiv V_{ij}$ (with $i$ and
$j$ nearest neighbors) are retained. The last assumption is valid only if
$W_r(r-r',\omega=0)$ decays rapidly in space.\footnote{This assumption is not necessarily valid
if low-energy metallic screening is due to particle-hole excitations within the 
low-energy window itself. However, extending the formalism from nearest-neighbor
to even longer range interactions is trivial, so we will not discuss this case explicitly.}  
The neglected non-site-diagonal parts of the electron-electron interactions
such as, for instance, the bond-charge-bond-charge matrix elements
$W\equiv V_{ijji}$, are believed to be small in usual solids.\cite{Campbell_1990}

\section{Derivation of the single-site EDMFT action using the cavity method \label{cavity_method}}

In the following we use the cavity method \cite{Georges_1996}
to derive the EDMFT action, Eq.~(\ref{eq:Effective_action_local_HS-UV}), and the EDMFT 
self-consistency equations which fix $\mathcal{G}$ and $\mathcal{U}$.
To this end, let us focus on a given site (denoted
by the index 0) and split the lattice action (Eq.~(\ref{eq:U-V_action_decoupled}))
into three parts: $S=S_{0}+S^{(0)}+\Delta S$ where $S_{0}$ denotes
the action of the site 0, $S^{(\text{0)}}$ the action of the lattice
with site 0 removed (the lattice with a {}``cavity'' at site 0)
and $\Delta S$ the remaining part:
\begin{widetext}
\begin{eqnarray}
S_{0} & = & \int_{0}^{\beta}d\tau\left\{ \sum_{\sigma}c_{0\sigma}^{*}\left(\partial_{\tau}-\mu\right)c_{0\sigma}+i\phi_{0}n_{0}+\frac{1}{2}\phi_{0}[v^{-1}]_{00}\phi_{0}\right\}, \\
\Delta S & = & \int_{0}^{\beta}d\tau\left\{ -\sum_{i\neq0, \sigma}t_{i0}\left(c_{0\sigma}^{*}c_{i\sigma}+c_{i\sigma}^{*}c_{0\sigma}\right)+\sum_{i}\phi_{i}[v^{-1}]_{i0}\phi_{0}\right\}, \\
S^{(0)} & = & \int_{0}^{\beta}d\tau\left\{ \sum_{ij\neq0,\sigma}c_{i\sigma}^{*}\left(\partial_{\tau}-\mu-t_{ij}\right)c_{j\sigma}+\frac{1}{2}\sum_{ij,\neq0}\phi_{i}[v^{-1}]_{ij}\phi_{j}+i\sum_{i\neq0}\phi_{i}n_{i}\right\}. 
\end{eqnarray}
\end{widetext}
Defining $\eta_{i\sigma}\equiv t_{i0}c_{0\sigma}$
and $j_{i}\equiv [v^{-1}]_{i0}\phi_{0}$, we can write $\Delta S=\int_{0}^{\beta}d\tau\left\{ -\sum_{i\ne 0,\text{\ensuremath{\sigma}}}(\eta_{i}^{*}c_{i\sigma}+c_{i\sigma}^{*}\eta_{i})+\sum_{i\ne 0}j_{i}\phi_{i}\right\} $, 
such that $\eta_{i\sigma}$ and $j_{i}$ can be regarded as sources
of correlation functions for the effective action of the site 0, defined
by $e^{-S_\text{eff}[c_{0}^{*},c_{0},\phi_{0}]}/Z_\text{eff}\equiv\int\mathcal{D}_{\neq0}[c_{i}^{*},c_{i},\phi_{i}]e^{-(S_{0}+S^{(0)}+\Delta S)}/Z$.
We can express the action as $S_\text{eff}=S_{0}-\Omega[\eta_{i}^{*},\eta_{i},j_{i}]+\mathrm{const}$, 
where $\Omega[\eta_{i}^{*},\eta_{i},j_{i}]\equiv\ln \int\mathcal{D}_{\neq0}[c_{i}^{*},c_{i},\phi_{i}]e^{-(S^{(0)}+\Delta S)}$
 is the generating functional of connected correlation functions
of the cavity,\cite{Negele_1988} 
\begin{widetext}
\begin{eqnarray}
G_{i_{1}\dots i_{n}j_{n}\dots j_{1}}^{(0)}(\tau_{1}\dots\tau_{n},\tau'_{1}\dots\tau'_{n}) & = & (-1)^{n}\frac{\delta^{2n}\Omega}{\delta\eta_{i_{1}}^{*}(\tau_{1})\dots\delta\eta_{i_{n}}^{*}(\tau_n)\delta\eta_{j_{n}}(\tau_1')\dots\delta\eta_{j_{1}}(\tau'_{n})},\\
W_{i_{1}\dots i_{n}j_{n}\dots j_{1}}^{(0)}(\tau_{1}\dots\tau_{n},\tau'_{1}\dots\tau'_{n}) & = & \frac{\delta^{2n}\Omega}{\delta j_{i_{1}}(\tau_{1})\dots\delta j_{i_{n}}(\tau_{n})\delta j_{j_{1}}(\tau_1')\dots\delta j_{j_{n}}(\tau'_{n})}.
\end{eqnarray}
An explicit expression for $\Omega$ is thus
\begin{eqnarray}
\Omega[\eta_{i}^{*},\eta_{i},j_{i}] & = & \sum_{n=1}^{\infty}\sum_{i_{1}\dots i_{n},j_{1}\dots j_{n}}\int d\tau_{1}\dots d\tau'_{n}\eta_{i_{1}}^{*}(\tau_{1})\dots\eta_{j_{1}}(\tau'_{n})(-1)^n G_{i_{1}\dots i_{n}j_{n}\dots j_{1}}^{(0)}(\tau_{1}\dots\tau'_{n})\label{eq:gen_functional}\nonumber\\
 & + & \sum_{n=1}^{\infty}\sum_{i_{1}\dots i_{n},j_{1}\dots j_{n}}\int d\tau_{1}\dots d\tau'_{n} j_{i_{1}}(\tau_{1})\dots j_{j_{n}}(\tau'_{n})W_{i_{1}\dots i_{n}j_{n}\dots j_{1}}^{(0)}(\tau_{1}\dots\tau'_{n}).
\end{eqnarray}
\end{widetext}

The DMFT approximation consists in approximating $\Omega$ by its
infinite-dimensional limit. In this limit, the hopping $t$ between
sites must be scaled as $t/\sqrt{z}$ (with $z=2d$) in order to keep
a finite kinetic energy, while $V$ must be scaled as $V/z$ in order to keep
the Hartree energy corresponding to the nearest-neighbor interaction
finite.\cite{Muller_1989} As a consequence of taking this limit,
all terms of order $n>1$ in Eq.~(\ref{eq:gen_functional}) vanish,
so that $\Omega^{DMFT}=\int d\tau d\tau'c_0^{*}(\tau)\left(-\sum_{ij}t_{i0}t_{j0}G_{ij}^{(0)}(\tau-\tau')\right)c_0(\tau')+\int d\tau d\tau'\phi_0(\tau)\left(\sum_{ij}v^{-1}_{i0}v^{-1}_{j0}W_{ij}^{(0)}(\tau-\tau')\right)\phi_0(\tau')$.
We thus arrive at the DMFT effective action of Eq.~(\ref{eq:Effective_action_local_HS-UV})
(where we have dropped the index 0 in order to simplify the notation)
with 
\begin{eqnarray}
\mathcal{G}^{-1}(i\omega_{n}) & \equiv & i\omega_{n}+\mu-\sum_{ij}t_{i0}t_{j0}G_{ij}^{(0)}(i\omega_{n}),\\
\mathcal{U}^{-1}(i\nu_{n}) & \equiv & v^{-1}_{00}-\sum_{ij} v^{-1}_{i0} v^{-1}_{j0}W_{ij}^{(0)}(i\nu_{n}).
\end{eqnarray}

Furthermore, in the limit of infinite dimensions, the cavity Green's
function is related to the lattice Green's function through $G_{ij}^{(0)}=G_{ij}-G_{i0}G_{0j}/G_{00}$
and $W_{ij}^{(0)}=W_{ij}-W_{i0}W_{0j}/W_{00}$, which is shown by
considering the paths contributing to $G_{ij}$ ($W_{ij}$) and not
to $G_{ij}^{(0)}$ ($W_{ij}^{(0)}$) (see Ref.~\onlinecite{Georges_1996} for more
details). This allows us to write $\sum_{ij}t_{i0}t_{j0}G_{ij}^{(0)}$, after Fourier-transforming, as
\begin{equation}
\sum_k \epsilon_k^2 G_k(i\omega_n)-\left(\sum_k\epsilon_k G_k(i\omega_n)\right)^2\bigg/\sum_k G_k(i\omega_n).
\label{eq:expr_cavity}
\end{equation}

 At this point a second approximation is made: the self-energies
are assumed to be $k$-independent, namely: $\Sigma(k,i\omega)\approx\Sigma_\text{loc}(i\omega)$
and $\Pi(k,i\nu)\approx\Pi_\text{loc}(i\nu)$. This also becomes exact
in the $d\rightarrow\infty$ limit.\cite{Metzner_1989} 
As a consequence, we can define the densities of states $\rho(\epsilon)=\sum_k \delta(\epsilon-\epsilon_k)$ and $\rho'(\epsilon)=\sum_k \delta(\epsilon-v_k^{-1})$, which allows us to rewrite (\ref{eq:expr_cavity}) as 
\begin{equation}
\int \frac{ d\epsilon \rho(\epsilon) \epsilon^2}{\zeta-\epsilon} - \left(\int \frac{ d\epsilon \rho(\epsilon) \epsilon}{\zeta-\epsilon}\right)^2 \bigg/\int \frac{ d\epsilon \rho(\epsilon)}{\zeta-\epsilon},
\end{equation}
where $\zeta \equiv i\omega_n+\mu-\Sigma_\text{loc}(i\omega_n)$. The same expression holds for the screened interaction, with $\rho\rightarrow \rho'$ and $\zeta\rightarrow \zeta'=[\tilde{v}^{-1}]_{00}-\Pi_\text{loc}(i\nu_n)$.

Using the following identities for Hilbert transforms:
\begin{eqnarray}
\int_{-\infty}^{\infty} \frac{d\epsilon \rho(\epsilon)\epsilon^2}{\zeta-\epsilon} &=& \zeta \int_{-\infty}^{\infty} \frac{d\epsilon \rho(\epsilon)\epsilon}{\zeta-\epsilon}\\
\int_{-\infty}^{\infty} \frac{d\epsilon \rho(\epsilon)\epsilon}{\zeta-\epsilon} &=& -1+\zeta \int_{-\infty}^{\infty} \frac{d\epsilon \rho(\epsilon)}{\zeta-\epsilon}
\end{eqnarray}
we obtain the self-consistency relations (\ref{eq:Self_consistency_G}) and (\ref{eq:Self_consistency_W}).

Equations (\ref{eq:Effective_action_local_HS-UV}), (\ref{eq:Self_consistency_G}) and (\ref{eq:Self_consistency_W})
form a closed set of equations: $S_\text{eff}^{DMFT}$, once solved, yields
$\Sigma_\text{loc}$ and $\Pi_\text{loc}$, which gives updated $\mathcal{G}$
and $\mathcal{U}$ which can in turn be used to solve the effective
local problem again until convergence is reached.

\section{Hamiltonian formulation of the impurity problem \label{hamiltonian_formulation}}

Some properties of action~(\ref{eq:Effective_action_2}) are more easily understood
in terms of its hamiltonian representation. The first two lines correspond to
an Anderson impurity model
\begin{equation}
H_{\mathrm{AIM}}=\sum_{p}\varepsilon_{p}a_{p}^{\dagger}a_{p}+\sum_{p}\left(V_{p}^{\sigma}a_{p\sigma}^{\dagger}c_{\sigma}+h.c.\right)+Un_{\uparrow}n_{\downarrow}-\mu n,
\label{eq:AIM}
\end{equation}
describing an impurity ($c$, $c^{\dagger}$) coupled to a bath of
non-interacting fermionic levels ($a_p$, $a^{\dagger}_p$, energy $
\varepsilon_p$). Here, $n_{\sigma} = c_\sigma^{\dagger} c_\sigma$ and $n = n_\uparrow + n_\downarrow$. The connection between Eqs.~(\ref{eq:Effective_action_2})
and (\ref{eq:AIM}) is given by $\mathcal{G}^{-1}(i\omega_{m})=i\omega_{m}+\mu-\Delta(i\omega_{m})$
and the hybridization function $\Delta(i\omega_{m})=\sum_{p}\frac{|V_{p}^{\sigma}|^{2}}{i\omega_{m}-\varepsilon_{p}}$.
On the other hand, the retarded effective interaction can be generated
by coupling the impurity to a bath of bosonic modes described by the Hamiltonian
\begin{eqnarray}
H_{\mathrm{boson}} & = & \sum_{p}\omega_{p}b_{p}^{\dagger}b_{p}+\sum_{p}\frac{\lambda_{p}}{\sqrt{2}}n(b_{p}+b_{p}^{\dagger})\nonumber\\
 & = & \sum_{p}\frac{\omega_{p}}{2}\left(\phi_{p}^{2}+\Pi_{p}^{2}\right)+\sum_{p}\lambda_{p}n_{0}\phi_{p},
\end{eqnarray}
with $\phi_{p}\equiv\frac{1}{\sqrt{2}}\left(b_{p}+b_{p}^{\dagger}\right)$
and $\Pi_{p}\equiv\frac{1}{i\sqrt{2}}\left(b_{p}-b_{p}^{\dagger}\right)$.

Using the identity $\Pi_p^2=-(\partial_{\tau}\phi_p(\tau))^2/\omega_p^2$, this can be written in an action formulation as
\begin{eqnarray}
S_{\mathrm{boson}}&=&\frac{1}{\beta}\sum_{m,p}\phi_{p}(i\nu_m)\left(\frac{-(i\nu_m)^{2}+\omega_{p}^{2}}{2\omega_{p}}\right)\phi_{p}(-i\nu_m)\nonumber\\
&+& \lambda_{p}\phi_{p}(i\nu_m)n(-i\nu_m)
\end{eqnarray}
Integrating out the bosonic degrees of freedom leads to
\begin{equation}
S_{\mathrm{boson}}=\frac{1}{\beta}\sum_{m}n(i\nu_{m})\left\{- \sum_{p}\lambda_{p}^{2}\frac{2\omega_{p}}{(i\nu_{m})^{2}-\omega_{p}^2}\right\} n(-i\nu_{m}).
\end{equation}
Defining $\mathcal{D}(i\nu_{m})=\int\frac{d\omega}{\pi}\mathrm{Im}\mathcal{D}(\omega)\frac{2\omega}{(i\nu_{m})^{2}-\omega^2}$
with 
\begin{equation}
\mathrm{Im}\mathcal{D}(\omega)\equiv - \pi\sum_{p}\lambda_{p}^{2}\delta(\omega-\omega_{p})\label{eq:Im_W_imp}
\end{equation}
 and Fourier-transforming $\mathcal{D}(i\nu_{n})$ yields the retarded interaction in Eq.~(\ref{eq:Effective_action_2}). 

The retarded interaction may thus be regarded as stemming from the
coupling to a bath of harmonic oscillators labeled by the index $p$, with frequency $\omega_{p}$
and coupling strength $\lambda_{p}$ (as already emphasized in Ref.~\onlinecite{Si_1996}). The effective interaction mediated by these auxiliary degrees of freedom is proportional to the squared coupling strength $\lambda_p^2$ times the free-phonon Green's function\cite{Mahan_1993} $D_{p}^{0}(i\nu_{n})\equiv - \langle T\phi_{p}(\tau)\phi_{p}(0)\rangle=\frac{2\omega_{p}}{(i\nu_{n})^{2}-\omega_{p}^{2}}$. Note that in complete analogy to the fermionic hybridization function $\Delta(\omega)$, the frequency dependent interaction
$\mathcal{D}(\omega)$ is determined self-consistently.


\end{document}